\documentclass[aps,prl,twocolumn,superscriptaddress,showpacs,amsmath,amssymb,floatfix]{revtex4-1}
\def\MyClass{0}
\def\MyFigSize{0.49\textwidth}
\def\MyFigSizeStar{0.99\textwidth}

\usepackage[T1]{fontenc}
\usepackage[utf8]{inputenc}
\usepackage[ngerman, english]{babel}
\usepackage{times}
\usepackage{amsmath} 
\usepackage{amsthm} 
\usepackage{amssymb} 
\usepackage{bm}
\usepackage{graphicx} 
\graphicspath{
{./figs/}%
}
\usepackage[yyyymmdd]{datetime}

\usepackage[usenames,dvipsnames,svgnames,table,rgb]{xcolor}
\usepackage{color}
\usepackage{setspace}
\usepackage{dcolumn}

\usepackage{longtable}

\newcommand{\eg}{\emph{e.g.\ }}
\newcommand{\ie}{\emph{i.e.\ }}

\addto\captionsenglish{}

\addto\captionsenglish{}

\newcommand{\mytitle}{Incorporation of Mn in Al$_{x}$Ga$_{1-x}$N probed by x-ray
  absorption and emission spectroscopy, high-resolution microscopy, x-ray
  diffraction and first-principles calculations}

\newcommand{\mydate}{\today}

\newcommand{\mypacs}{81.05.Ea, 61.05.cj, 78.70.En, 68.55.Ln}

\usepackage{hyperref}
\hypersetup{%
  colorlinks=true,
  linkcolor=blue,
  urlcolor=blue,
  citecolor=blue,
  pdftitle={\mytitle},%
  pdfauthor={M. Rovezzi et al.},%
  pdfsubject={},%
  pdfkeywords={}%
}

\begin{document}

\ifnum\MyClass=0
\title{\mytitle}
\author{Mauro~Rovezzi}\email{mauro.rovezzi@esrf.eu}
\affiliation{European Synchrotron Radiation Facility, 71 avenue des Martyrs, CS 40220 Grenoble, France}
\author{Wolfgang~Schl\"ogelhofer}
\affiliation{European Synchrotron Radiation Facility, 71 avenue des Martyrs, CS 40220 Grenoble, France}
\affiliation{Institute of Semiconductor and Solid State Physics, Johannes Kepler University, Altenberger Str. 69, A-4040 Linz, Austria}
\author{Thibaut~Devillers}
\affiliation{Institute of Semiconductor and Solid State Physics, Johannes Kepler University, Altenberger Str. 69, A-4040 Linz, Austria}
\author{Nevill~Gonzalez~Szwacki}
\affiliation{Institute of Theoretical Physics, Faculty of Physics, University of Warsaw, ul. Pasteura 5, PL-02-093 Warszawa, Poland}
\author{Tian~Li}
\affiliation{Institute of Physics, Polish Academy of Sciences, Al.\,Lotnik\'ow 32/46, PL-02-668 Warszawa, Poland}
\author{Rajdeep~Adhikari}
\affiliation{Institute of Semiconductor and Solid State Physics, Johannes Kepler University, Altenberger Str. 69, A-4040 Linz, Austria}
\author{Pieter~Glatzel}
\affiliation{European Synchrotron Radiation Facility, 71 avenue des Martyrs, CS 40220 Grenoble, France}
\author{Alberta~Bonanni}\email{alberta.bonanni@jku.at}
\affiliation{Institute of Semiconductor and Solid State Physics, Johannes Kepler University, Altenberger Str. 69, A-4040 Linz, Austria}
\date{\mydate}
\pacs{\mypacs}
\fi

\begin{abstract}
  Synchrotron radiation x-ray absorption and emission spectroscopy techniques,
  complemented by high-resolution transmission electron microscopy methods and
  density functional theory calculations are employed to investigate the effect
  of Mn in Al$_x$Ga$_{1-x}$N:Mn samples with an Al content up to 100\%. The atomic and
  electronic structure of Mn is established together with its local environment
  and valence state. A dilute alloy without precipitation is obtained for
  Al$_x$Ga$_{1-x}$N:Mn with Al concentrations up to 82\%, and the surfactant role of Mn in
  the epitaxial process is confirmed.\\
\end{abstract}

\ifnum\MyClass=0
\maketitle
\fi

\section{Introduction}
\label{sec:introduction}
%
Hetero-structures based on III-nitrides \cite{Morkoc:2008_book} and in
particular on the combination Al$_x$Ga$_{1-x}$N/GaN represent the basis of a variety of
state-of-the-art (opto)electronic devices like blue and white light-emitting
diodes \cite{Gutt:2012_APEX}, laser diodes \cite{Yoshida:2008_NP}, blue lasers
\cite{Nakamura:2000_book}, high-power- \cite{Shur:1998_SSE}, and
high-electron-mobility-transistors \cite{Mishra:2002_conf}. Most of the above
mentioned devices are commercially available and their performance continuously
improved. Furthermore, III-nitrides doped with transition metals (TM) have also
been the focus of considerable research efforts towards the demonstration of
semiconductor spintronic functionalities \cite{Dietl:2014_RMP}. In this respect,
while a remarkable number of reports on GaN:Mn provide an overview on the
structural, optical, magnetic and electric properties of this material system
\cite{Kondo:2002_JCG, Martinez-Criado:2005_APL, Sarigiannidou:2006_PRB,
  Freeman:2007_PRB,Bonanni:2011_PRB, Sawicki:2012_PRB, Stefanowicz:2013_PRB,
  Kunert:2012_APL}, little is known about Al$_x$Ga$_{1-x}$N:Mn
\cite{Polyakov:2004_SST,Polyakov:2004_JEM,Reed:2011_conf,Frazier:2003_JAP} and
related nanostructures \cite{Seong:2006_AdvMat}. Recent findings
\cite{Devillers:2015_CGD} indicate this alloy as particularly interesting for
\eg the self-assembling of functional multilayers and for having revealed the
decisive role of Mn as surfactant during the epitaxial growth of Al$_x$Ga$_{1-x}$N:Mn,
considerably enhancing the critical thickness of Al$_x$Ga$_{1-x}$N:Mn on GaN, and opening
new perspectives for the realization of \eg improved reflectors in GaN-based
laser structures. We report here on Al$_x$Ga$_{1-x}$N:Mn grown by means of metalorganic
vapor phase epitaxy (MOVPE) in a broad range of Al concentrations and
extensively investigated $via$ x-ray absorption spectroscopy (XAS), x-ray
emission spectroscopy (XES), energy-dispersive spectrometry (EDS), x-ray
diffraction (XRD), and high-resolution (HR) transmission electron microscopy
(TEM), supported by density functional theory (DFT) calculations. The results
provide fundamental information on the microstructure and local environment in
the layers and on the valence state of Mn incorporated in the lattice over the
whole range of Al concentrations.\\

%
\section{Experimental and theoretical methods}
\label{sec:methods}
%
The wurtzite (wz) Al$_x$Ga$_{1-x}$N:Mn samples are grown in an AIXTRON 200RF
horizontal-tube MOVPE reactor. All structures are deposited on $c$-plane
sapphire substrates with trimethylgallium (TMGa), trimethylaluminum (TMAl),
bis-methylcyclopentadienyl-manganese (MeCp$_2$Mn) and ammonia (NH$_3$) as
precursors for respectively Ga, Al, Mn, N, and with H$_2$ as carrier gas. The
epitaxial process, developed from a well established procedure
\cite{Bonanni:2003_JCG}, consists of: (i) substrate nitridation; (ii) low
temperature (540 $^{\circ}$C) deposition of a GaN nucleation layer (NL); (iii)
its annealing under NH$_3$; (iv) growth of a 1 $\mu$m device-quality GaN buffer
deposited at 1020 $^{\circ}$C; (v) Al$_x$Ga$_{1-x}$N:Mn layers at 850
$^{\circ}$C, with the same TMGa and MeCp$_2$Mn flow rates and different - over
the sample series - TMAl flow rates ranging from 1 to 80 standard cubic
centimeters per minute (sccm). In order to have real time control over the
entire fabrication process, the MOVPE system is equipped with an \textit{in
  situ} Isa Jobin Yvon ellipsometer that allows for both spectroscopic and
kinetic measurements in the energy range 1.5 eV -- 5.5\,eV
\cite{Peters:2000_JAP}. The structures are routinely characterized by atomic
force microscopy (AFM), secondary-ion mass spectroscopy (SIMS) and
(magneto)photoluminescence (PL) in order to get information on the surface
roughness, chemical composition and magnetooptical response,
respectively. Measurements of SQUID magnetometry in the temperature range
between 1.5 K and room temperature, confirm the samples to be
paramagnetic. Here, we focus on the effect of Mn incorporation on the structural
arrangement of Al$_x$Ga$_{1-x}$N:Mn and on the local atomic environment of Mn,
with particular attention to the XRD and HRTEM analysis as essential complement
to the synchrotron XAS and XES measurements. All considered Al$_x$Ga$_{1-x}$N:Mn
samples are listed together with their growth parameters in
Table~\ref{tab:growth}. The Mn concentration in all doped layers is $\approx$1\%
cations, as established by SIMS analysis.\\
\begin{table}[!htb]
  \caption{Growth parameters for the Al$_x$Ga$_{1-x}$N:Mn samples presented in this
    work. Al concentration $x$ (from XRD); TMGa and TMAl flow rates and the
    pressure $P$ in the reactor during the process. The MeCp$_2$Mn and NH$_3$
    flow rates are fixed at 490\,sccm and 1500\,sccm, respectively; the
    substrate temperature during the growth of the GaN buffer layer and during
    the deposition of the Al$_x$Ga$_{1-x}$N:Mn layer are, respectively, 1020\,$^\circ$C and
    850\,$^\circ$C. The nominal thickness is obtained from the kinetic
    ellipsometry spectra and confirmed by TEM cross-sections.}
  \label{tab:growth}
  \begin{ruledtabular}
    \begin{tabular}{c|ccccc}
      sample  & $x$ & TMGa & TMAl & $P$  & thickness\\           
              & \%  & sccm & sccm & mbar & nm \\
      \hline
      \#A        &   0   &  1  &   0  &  200 & 500 \\ 
      \#B	 &   12  &  1  &   1  &  100 & 260 \\ 
      \#C	 &   20  &  1  &   3  &  100 & 293 \\ 
      \#D	 &   41  &  1  &   9  &  100 & 377 \\ 
      \#E	 &   59  &  1  &  27  &  100 & 553 \\ 
      \#F	 &   71  &  1  &  80  &  100 & 845 \\ 
      \#G	 &   82  &  1  &  80  &   50 & 780 \\ 
      \#H	 &  100  &  0  &  80  &  100 & 553 \\ 
    \end{tabular}
  \end{ruledtabular}
\end{table}
High resolution XRD measurements are carried out in a PANalytical's X'Pert PRO
Materials Research Diffractometer (MRD) equipped with a hybrid monochromator
(parabolic-shaped multilayer mirror and a channel-cut Ge crystal) and a
1/4$^{\circ}$ divergence slit. The diffracted beam is measured with a
solid-state PixCel detector used as 256-channels detector with a 11.9\,mm
anti-scatter slit. For the whole series of Al$_x$Ga$_{1-x}$N:Mn samples, $\theta$-2$\theta$
scans are acquired for $2\theta$ values between 30$^\circ$ and 80$^\circ$ and
complemented with maps of asymmetric diffraction peaks. These measurements
provide information on the composition and strain state of the films
\cite{Moram:2009_RPP}.\\
%
Cross-sectional TEM specimen are prepared by mechanical polishing, dimpling and
final ion milling in a Gatan Precision Ion Polishing System. The samples are
studied using both conventional and scanning transmission electron microscopy
(CTEM/STEM) for bright/dark-field (BF/DF), HRTEM and high angle annular dark
field (HAADF) imaging. The energy dispersive x-ray spectrometry (EDS) technique
is employed to analyze the chemical distribution of the various elements in the
samples. The measurements reported here are performed in a FEI Titan Cube 80-300
operating at 300\,keV, while a JEOL 2010F operating at 200\,keV is routinely
employed for preliminary characterization of all the grown samples.\\
%
The x-ray absorption and emission measurements at the Mn K-edge (6539\,eV) are
carried out at the beamline ID26 at the European Synchrotron Radiation Facility
(ESRF). The incoming x-ray beam, linearly polarized in the horizontal plane, is
produced by three coupled undulators (u35) and monochromatized using a
cryogenically cooled double Si(111) crystal monochromator. Harmonics rejection
and heat load removal are achieved by using three Si mirrors at glancing angle
of 2.5\,mrad. The beam focusing (horizontal and vertical) is performed by means
of two Si bent mirrors. This configuration permits to obtain a beam size of
$\approx$\,(600$\times$100)\,$\mu$m$^2$ (horizontal $\times$ vertical) and a flux
of $\approx$10$^{13}$ ph/s on the sample. The measurements are carried out in
fluorescence mode at room temperature and under nitrogen flow to avoid
depositing ambient impurities on the samples' surface. The total fluorescence
yield (TFY) spectra are obtained with a Si photodiode, while the high energy
resolution fluorescence detected (HERFD) spectra are acquired with a wavelength
dispersive spectrometer equipped with 5 spherically bent crystal analyzers
(bending radius of 1 m) and an avalanche photodiode arranged in a vertical
point-to-point Rowland circle geometry \cite{Glatzel:2012_JESRP}. The HERFD-XAS
data are collected at the maximum of the K$\alpha_1$ emission line using Ge(333)
analyzers. The XES measurements are performed at the K$\beta$ core-to-core lines
(K$\beta^{\prime}$ and K$\beta_{1,3}$) using Si(440) analyzers and with the
incoming excitation set at 6700\,eV. For these configurations, the total energy
resolutions (convolution of monochromator and spectrometer) are, respectively,
$\approx$1.3 eV and $\approx$1.0 eV (full-width-at-half-maximum). In addition, to
exploit the natural linear x-ray dichroism (XLD) arising from the wurtzite
hexagonal lattice \cite{Brouder:1990_JPCM}, two geometries are employed: the
vertical grazing incidence (VGI) and the horizontal grazing incidence (HGI). The
grazing angle fixed at $\approx$5$^\circ$ permits to approximate the two
configurations, respectively, to $\epsilon \parallel c$ and $\epsilon \perp c$,
where $\epsilon$ is the polarization vector and $c$ is the wurtzite $c$-axis
that corresponds to the sample's surface normal. The number of acquired spectra
and the integration time per energy point are chosen in order to reach an edge
jump of $\approx$10$^6$ total counts per spectrum on each specimen. This permits to
obtain the same stastistics for all samples. The HERFD- and TFY-mode spectra are
collected in the near-edge and extended regions (XANES and EXAFS) for the whole
series.\\
%
Theoretical calculations are performed to support the analysis of the
experimental XANES and EXAFS data. In order to simulate the Al$_x$Ga$_{1-x}$N:Mn
series, seven wurtzite supercells (SC), $3a\times3b\times2c$ (72 atoms), are
built using the program VESTA \cite{Momma:2011_JAC}, with Al concentrations
corresponding to those found experimentally, as reported in
Table~\ref{tab:growth}. The experimental lattice parameters established from XRD
measurements are employed for the SC, while the wurtzite $u$ parameter is chosen
to the average value of $u_{\rm avg}$\,=\,0.38 from
Ref.~\cite{Paszkowicz:2004_JAC}. To simulate the Mn incorporation in the
Al$_x$Ga$_{1-x}$N lattice the following defect configurations are taken into
account for one Mn atom as: 1) substitutional of Ga or Al (Mn$_{\rm S}$); 2)
interstitial in the tetrahedral (Mn$_{\rm IT}$) or octrahedral (Mn$_{\rm IO}$)
sites with Wyckoff positions, (2/3, 1/3, $u$/2) and (0, 0, $u$/2),
respectively. This corresponds to a Mn concentration of $\approx$1\%.\\
%
The lattice parameters and atomic positions of the SC are additionally relaxed
by means of DFT using the {\sc quantum-espresso} package
\cite{Giannozzi:2009_JPCM}. The first-principles spin-polarized calculations are
performed using a plane-wave basis and the projector augmented wave (PAW) method
\cite{Blochl:1994_PRB}. The exchange correlation energy is described by the
Perdew-Burke-Ernzerhof parametrization within the generalized gradient
approximation (PBE-GGA) \cite{Perdew:1996_PRL}. The Hubbard correction
(DFT-GGA+U framework) is applied to Mn with U parameter equal to 3.9 eV
\cite{GonzalezSzwacki:2011_PRB}. The plane-waves cutoff energy is set at 60\,Ry
to ensure convergence and the irreducible Brillouin zone is sampled with the
Monkhorst-Pack scheme \cite{Monkhorst:1976_PRB} using a 4$\times$4$\times$4
$k$-point mesh. For each Al concentration ($x$), the formation energies of Mn
impurities substituting Ga or Al (Mn$_{\rm Ga, Al}$) in Al$_x$Ga$_{1-x}$N (AlGaN) are
calculated through E$_{\rm f}$[Mn$_{\rm Ga,Al}$] = E[Mn$_{\rm Ga,Al}$] +
E[AlGaN] - $\mu_{\rm Mn}$ + $\mu_{\rm Ga,Al}$, where E[Mn$_{\rm Ga,Al}]$ and
E[AlGaN] are the total energies of Al$_x$Ga$_{1-x}$N:Mn and undoped Al$_x$Ga$_{1-x}$N,
respectively. $\mu_{\rm Mn}$ and $\mu_{Ga,Al}$ are the atom chemical potentials
obtained from bulk $\alpha$-Mn, $\alpha$-Ga and Al.\\
%
The Mn K-edge XANES and EXAFS spectra are simulated within the real-space
Green's function formalism employing the {\sc fdmnes} \cite{Bunau:2009_JPCM} and
{\sc feff9} \cite{Rehr:2010_PCCP} codes, respectively. The muffin-tin potentials
and the Hedin-Lunqvist approximation \cite{Hedin:1971_JPC} for the
exchange-correlation component are used. The calculations are performed using
the DFT-relaxed SC, rescaled to the experimental lattice parameters as input
structures. The cluster radius for the spectra is set to 10\,\AA, while the
self-consistent field (SCF) loop is swept within a radius of 6\,\AA. For the
comparison with the experiment, the XANES spectra are consequently convoluted
with a Lorentzian function with an energy-dependent arctangent-like width,
$\Gamma(E)$ \cite{Bunau:2009_JPCM}. This model correctly accounts for the
core-hole and the photo-electron mean-free-path broadening. The best agreement
with the experimental data is found going from $\Gamma_{\rm min}$\,=\,0.5\,eV to
$\Gamma_{\rm max}$\,=\,4.0\,eV. A second convolution with a Gaussian function of
constant width (0.9\,eV) is also applied to take into account the experimental
broadening. These parameters, below the core-hole lifetime, are in line with the
expected sharpening effect due to the high resolution detection
\cite{Rovezzi:2014_SST}. The EXAFS signal is extracted from the absorption
spectra $via$ the {\sc viper} code \cite{Klementev:2001_JPDAP}, using a
smoothing spline algorithm and selecting the edge energy $E_0$ at the maximum of
the derivative peak corresponding to the typical shoulder after the pre-edge
features. The EXAFS quantitative analysis, that is based on scattering paths
expansion, Fourier transform and least-squares fits, is performed with the {\sc
  ifeffit} \cite{Newville:2001_JSR,Ravel:2005_JSR} software. The EXAFS
Debye-Waller factors (DWF) for the multiple scattering paths are modeled as the
sum of the DWF of single scattering paths plus a Debye model with room
temperature (300 K) target and a Debye temperature of 600 K
\cite{Passler:2007_JAP}. In both XANES and EXAFS simulations, the polarization
effects \cite{Brouder:1990_JPCM} are correctly included.\\
%
\section{Results and discussion}
\label{sec:results}
%
\begin{figure}[!htb]
  \centering
  \includegraphics[width=\MyFigSize]{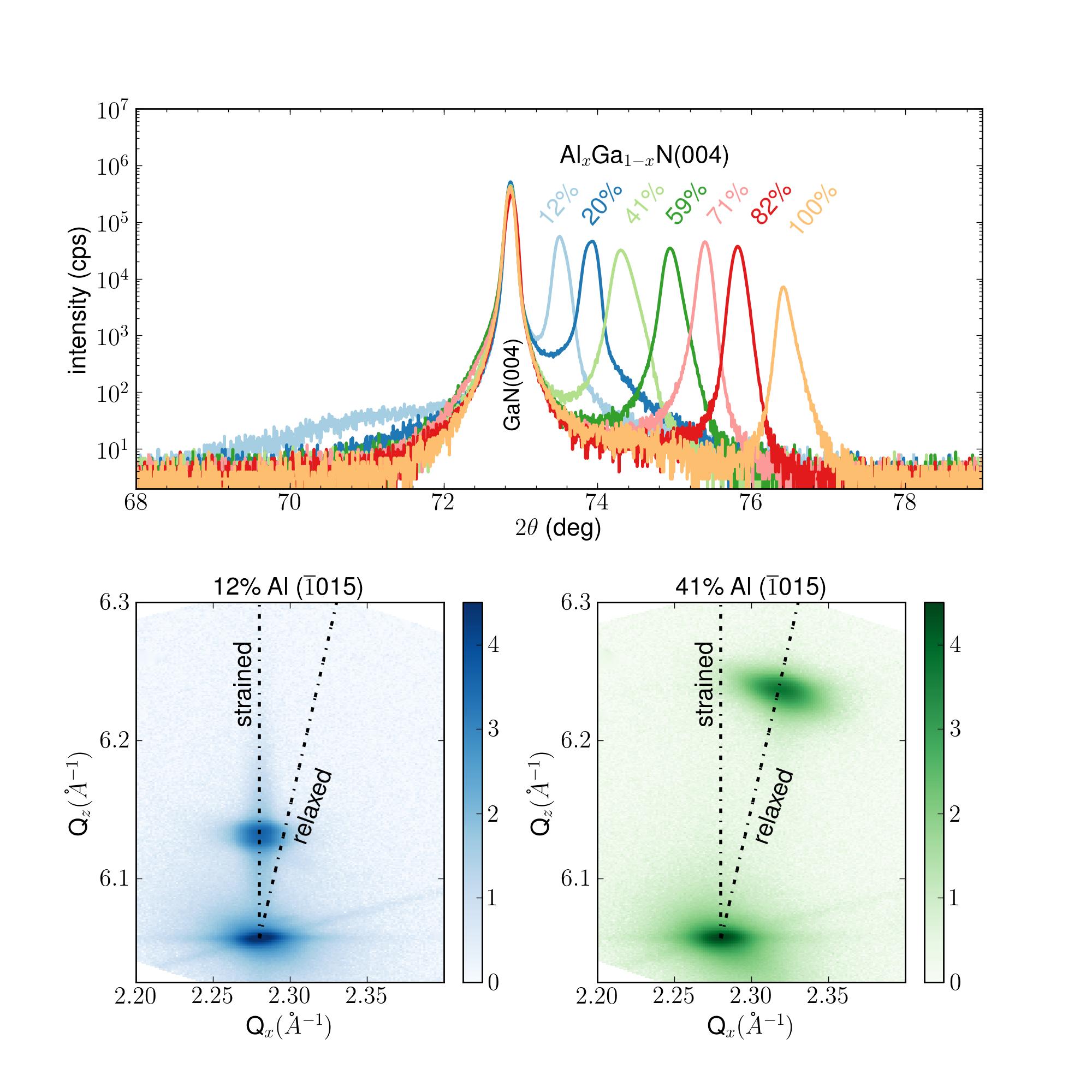}
  \caption{(Color online) XRD: (top panel) evolution of the GaN and Al$_x$Ga$_{1-x}$N (004)
    peak position over the whole series; (bottom panels) maps of the
    ($\overline{1}$015) asymmetric reflection of GaN and Al$_x$Ga$_{1-x}$N measured for
    Al$_x$Ga$_{1-x}$N:Mn with 12\% and 41\% Al, respectively.}
  \label{fig:xrd}
\end{figure}
\begin{figure*}[!htb]
  \centering
  \includegraphics[width=\MyFigSizeStar]{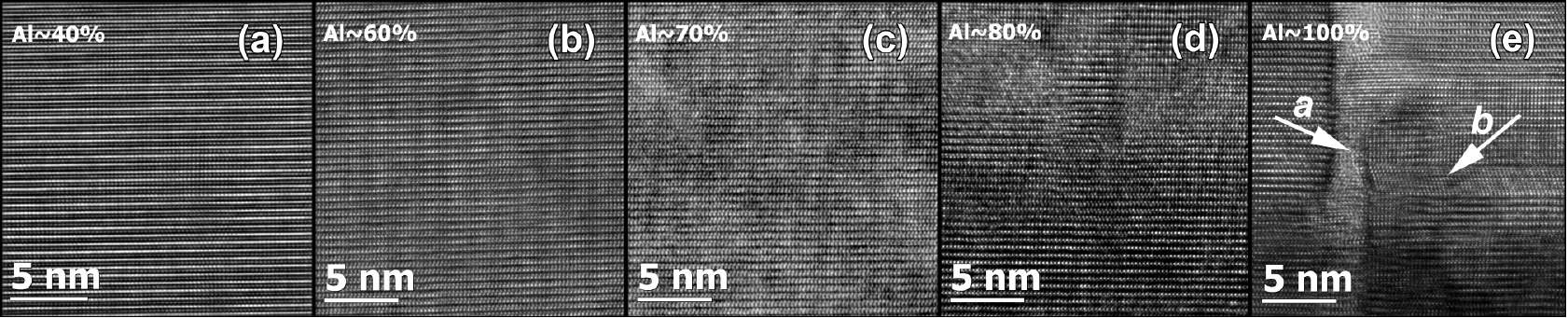}
  \caption{HRTEM images of Al$_x$Ga$_{1-x}$N:Mn layers (a) 41\%, (b) 59\%, (c)
    71\%, (d) 82\% and (e) 100\% (samples \#D, \#E, \#F, \#G and \#H in
    Table~\ref{tab:growth}).}
  \label{fig:hrtem} 
\end{figure*}
\begin{figure}[!htb]
  \centering
  \includegraphics[width=\MyFigSize]{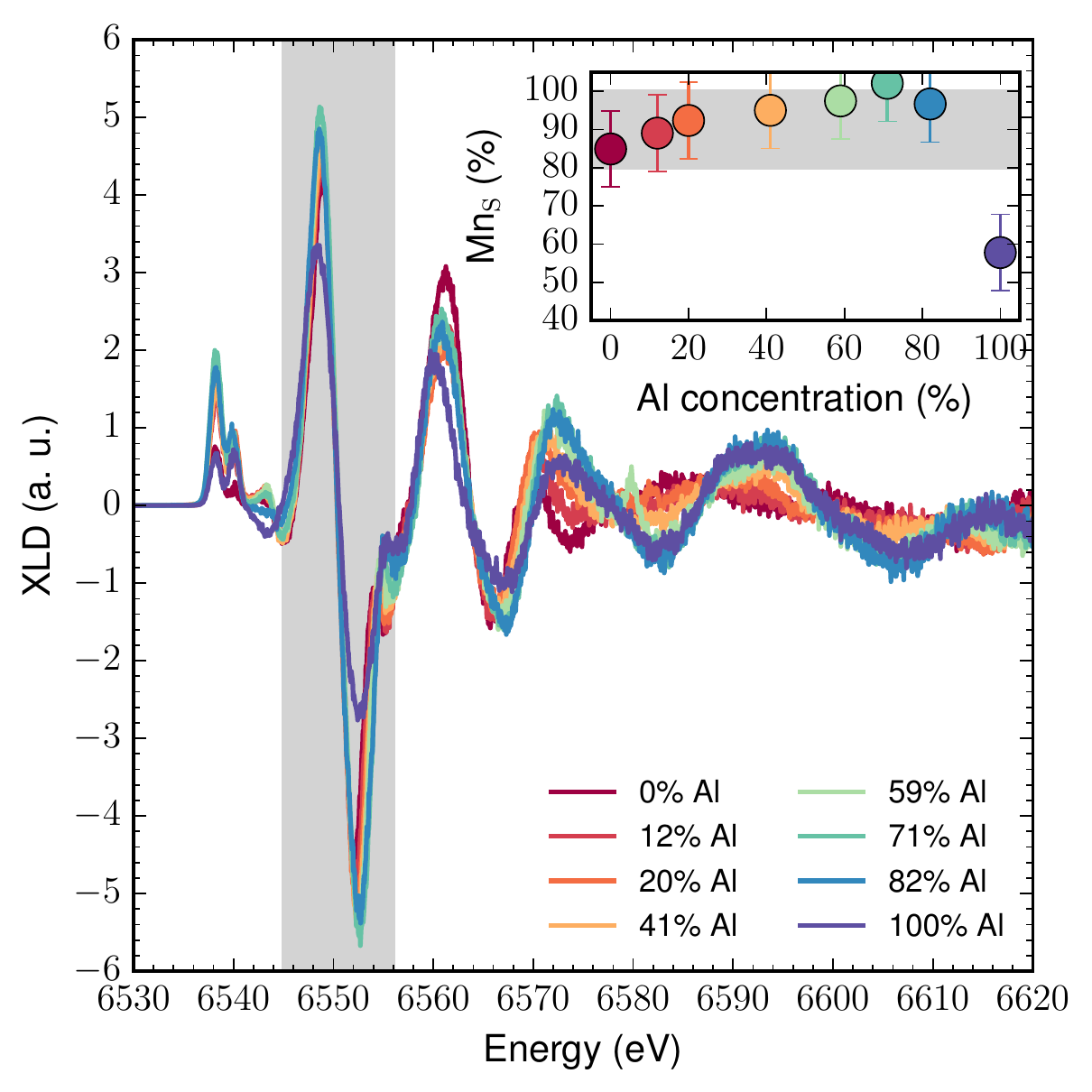}
  \caption{(Color online) XLD signal for the Al$_x$Ga$_{1-x}$N:Mn series. The amplitude in
    the highlighted region is taken as a figure of merit for Mn$_{\rm
      S}$. Inset: quantitative analysis of the results.}
  \label{fig:xld}
\end{figure}
\begin{figure}[!htb]
  \centering
  \includegraphics[width=\MyFigSize]{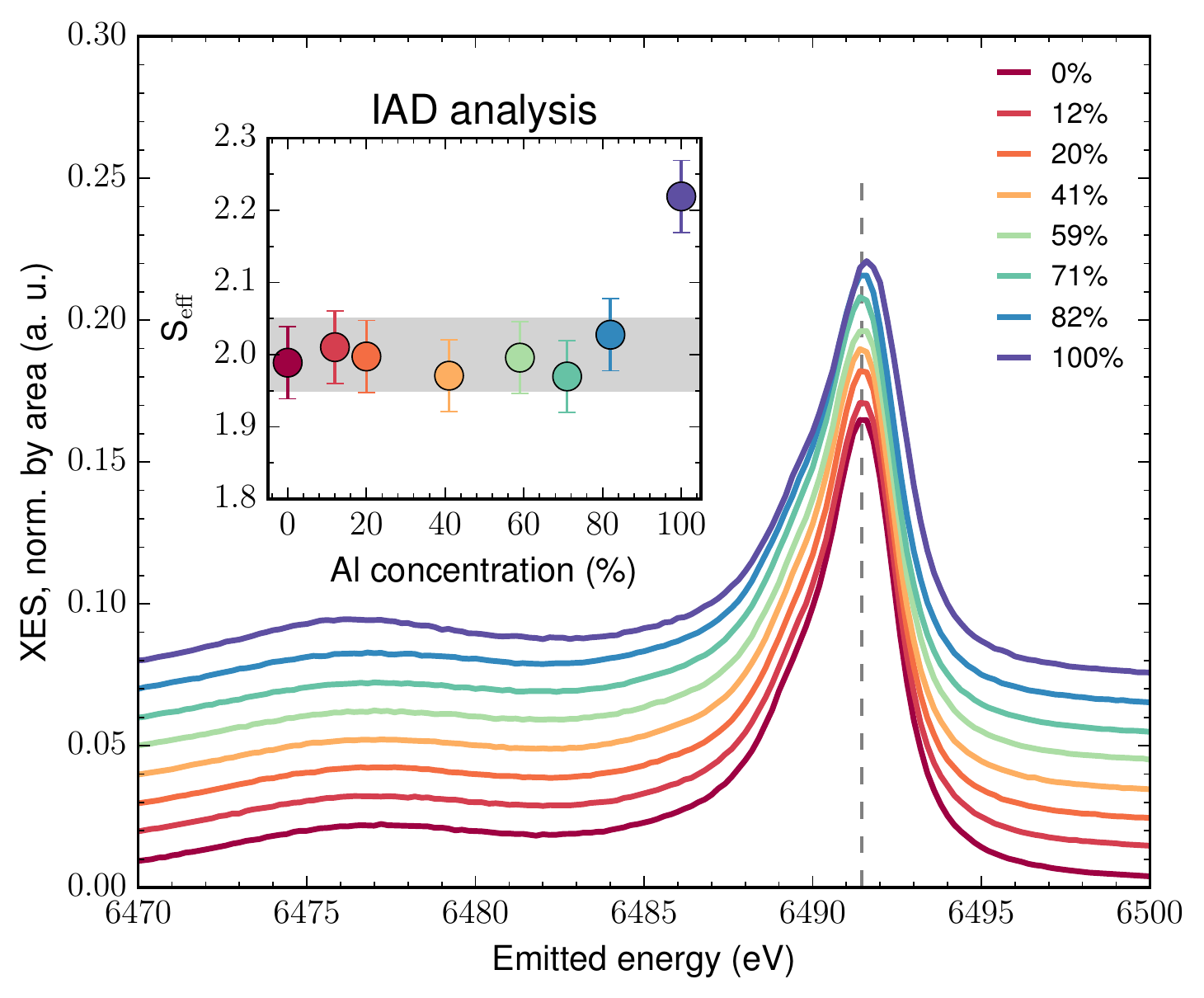}
  \caption{(Color online) Mn K$\beta$ XES data over the whole series of
    samples. Inset: relative IAD analysis. The vertical dashed line is a guide
    to the eye.}
  \label{fig:xes}
\end{figure}

As a first step, we determine the Al content from the strain analysis on the XRD
data. The XRD spectra of the symmetric (004) reflection over the whole series
are reported in the top panel of Fig.~\ref{fig:xrd}. In the $\theta$-2$\theta$
scans for $2\theta$ values between 30$^\circ$ and 80$^\circ$ on all considered
samples only reflections from the sapphire substrate (not shown), from the GaN
buffer, and from the Al$_x$Ga$_{1-x}$N layers are detectable, with no indication
of secondary phases. From the position of the Al$_x$Ga$_{1-x}$N\ peak it is
possible to deduce the Al$_x$Ga$_{1-x}$N \emph{c}-parameter. In order to gain
insight into the Al content in the films, maps of the ($\overline{1}$015)
asymmetric reflection have been acquired for the whole series and are shown in
the bottom panels of Fig.~\ref{fig:xrd} for the films containing 12\% and 40\%
of Al. The strain state of the Al$_x$Ga$_{1-x}$N layer is deduced from the
relative position of the ($\overline{1}$015) reflection of GaN and
Al$_x$Ga$_{1-x}$N, and the \emph{a} and \emph{c} lattice parameters are obtained
from the $Q_{x}$ and $Q_{z}$ coordinates of the Al$_x$Ga$_{1-x}$N
($\overline{1}$015) reflection, upon a 2D Gaussian fit. To extract the Al
concentration, we consider a linear variation of the lattice parameters between
GaN and Al$_x$Ga$_{1-x}$N as a function of the Al concentration according to the
Vegard's law \cite{Denton:1991_PRA} for the relaxed structures. For the strained
samples, the compressibility of Al$_x$Ga$_{1-x}$N through the Poisson
coefficient is taken into account. It is important to remark that in the set of
samples studied, the layer is either fully strained (\#A to \#C) or fully
relaxed (\#D to \#H) where the full relaxation is likely to be due to cracks
crossing the layer down to the interface with GaN. The Al concentrations
obtained from XRD -- as summarized in Table~\ref{tab:growth} -- are coherent
within 1\% error with those measured by EDS. The Al$_x$Ga$_{1-x}$N experimental
lattice parameters are reported in Table~\ref{tab:lattice}. The computed lattice
parameters closely follow the Vegard's law in accord with previous works based
on full-potential augmented plane wave method calculations
\cite{Dridi:2003_SST}. Nevertheless, the computed lattice parameters
overestimate the experimental values by $\approx$1\%. This is explained by the
strong dependence of DFT on the level of theory employed. For this reason, we
force in the DFT-relaxed supercells the experimental lattice parameters.\\
\begin{table}[!htb]
  \caption{Lattice parameters and strain state found experimentally with
    XRD. The error bar on the last digit is reported in parenthesis. It
    corresponds to the error propagation from the full-width-at-half-maximum
    ($\approx$2.35 $\sigma$) of the fitted two-dimensional Gaussian peak, that
    is, $\pm 0.01 \AA$ and $\pm 0.004 \AA$ for $a$ and $c$, respectively.}
  \label{tab:lattice}
  \begin{ruledtabular}
    \begin{tabular}{c|cccc}
      sample  & $x$ & $a$ & $c$ & strain\\           
              & \%  & \AA & \AA & state \\
      \hline
      \#A  &   0   &  3.18(1)  & 5.187(4) & strained \\ 
      \#B  &   12  &  3.18(1)  & 5.148(4) & strained \\ 
      \#C  &   20  &  3.18(1)  & 5.123(4) & strained \\ 
      \#D  &   41  &  3.16(1)  & 5.100(4) & relaxed  \\ 
      \#E  &   59  &  3.14(1)  & 5.063(4) & relaxed  \\ 
      \#F  &   71  &  3.13(1)  & 5.038(4) & relaxed  \\ 
      \#G  &   82  &  3.12(1)  & 5.014(4) & relaxed  \\ 
      \#H  &  100  &  3.11(1)  & 4.980(4) & relaxed  \\ 
    \end{tabular}
  \end{ruledtabular}
\end{table}
%
We investigate $via$ DFT also the formation energies upon relaxation for the
incorporation of Mn in Al$_x$Ga$_{1-x}$N. First of all, we study the total
energies of the Al$_x$Ga$_{1-x}$N\ alloy without Mn with respect to atomic-scale
composition fluctuations. This permits to understand whether the alloy behaves
locally as an ordering of GaN and AlN separate unit cells or there is a random
distribution of Al/Ga atoms among the cation positions in the supercell. For the
intermediate Al concentrations $x$ = 0.25, 0.5, and 0.75, we compare the total
energies for several structures with random position of Al and Ga atoms in
cation sites. This means that, for a given intermediate Al concentration - that
is, for a given number of Al atoms in the supercell - we randomly change the
position of Al/Ga atoms in cation sites and calculate the total energy of each
configuration. We find that the total energies of those configurations for each
Al concentration do not differ by more than 50 meV. From this result, we
conclude that the Al$_x$Ga$_{1-x}$N\ alloy has Al and Ga cations in random
positions for all concentrations of the constituents. The second step consists
in investigating the formation energies upon incorporation of Mn at
substitutional and interstitial sites. For the substitutional site, we assume
that for $x \le 0.5$ Mn substitutes mostly Ga sites, whereas for $x > 0.5$ Mn
ions replace Al positions. It is found that Mn$_{\rm S}^{\rm Ga}$ ($x \le 0.5$)
has a constant formation energy of 3.5 eV, while for Mn$_{\rm S}^{\rm Al}$ ($x >
0.5$) an abrupt increase in the formation energy to 5.5 eV is obtained. This
result indicates that, in terms of formation energy, Mn tends to substitute Ga
atoms rather than Al ones, challenging the epitaxy of high-quality AlN:Mn. On
the other hand, this does not take into account the surface energies that play a
crucial role during growth. For the interstitial sites (tetrahedral, Mn$_{\rm
  IT}$ and octahedral, Mn$_{\rm IO}$), we find always formation energies higher
than the one of Mn$_{\rm S}$. Upon relaxation, Mn$_{\rm IO}$ remains at its
nominal site, with a formation energy increasing linearly with $x$, from 6.5 eV
($x=0$) to 9.25 eV ($x=1$). On the other hand, Mn$_{\rm IT}$ is rather unstable
and tends to move toward Mn$_{\rm IO}$; its formation energy is $\approx$8.25
eV, regardless of $x$. These results show that the substitutional incorporation
of Mn in Al$_x$Ga$_{1-x}$N\ is favored. As confirmed by the experimental data
reported in the following.\\
\begin{figure*}[!htb]
  \centering
  \includegraphics[width=\MyFigSizeStar]{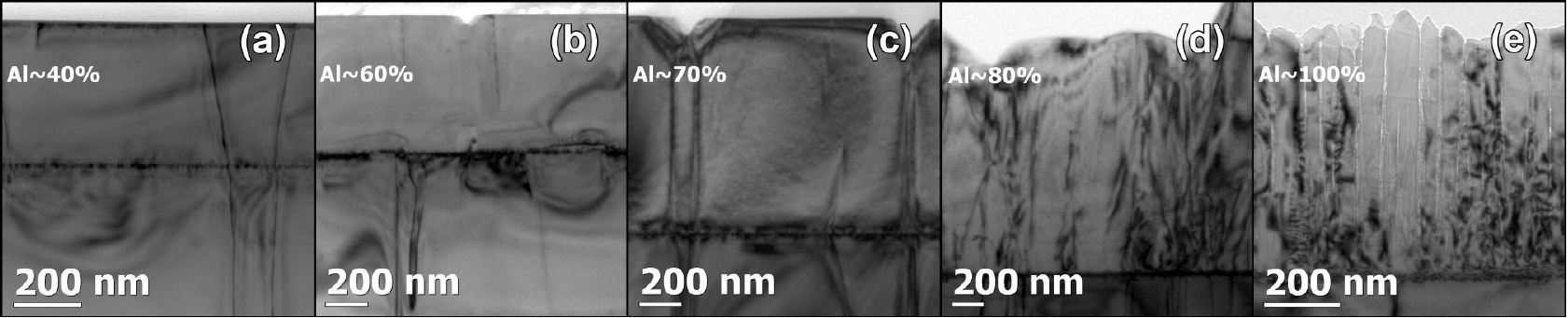}
  \caption{TEM micrographs of Al$_x$Ga$_{1-x}$N:Mn layers with (a) 41\%, (b)
    59\%, (c) 71\%, (d) 82\% and (e) 100\% (samples \#D, \#E, \#F, \#G and \#H
    in Table~\ref{tab:growth}).}
  \label{fig:tem} 
\end{figure*} 
According to the TEM micrographs shown in Fig.~\ref{fig:tem}, the layers are
structurally homogeneous for Al concentrations up to 82\%. Moreover, EDS spot
sampling and line scans (not shown) confirm that the layers are chemically
homogeneous. In contrast to the layer-by-layer growth of Al$_x$Ga$_{1-x}$N:Mn up
to Al concentrations as high as 82\% shown in Figs.~\ref{fig:tem}(a)--(d), the
columnar structure of the AlN:Mn sample is evidenced in
Fig.~\ref{fig:tem}(e). The Al$_x$Ga$_{1-x}$N:Mn layer with 82\% Al is still
structurally coherent with the GaN buffer layer, but at the boundary between 2D
and 3D growth. The homogeneous structure of the Al$_x$Ga$_{1-x}$N:Mn layers with
Al (a) 41\%, (b) 59\%, (c) 71\% and (d) 82\% is evidenced by the HRTEM images
taken close to the [11$\overline{2}$0] zone axis and reported in
Fig.~\ref{fig:hrtem}. According to a Fast Fourier Transform (FFT) analysis,
there is no compositional ordering or modulation of the Al concentration, in
contrast to what reported previously for Al$_x$Ga$_{1-x}$N layers without Mn
\cite{Gao:2006_JAP}. In the HRTEM image of Fig.~\ref{fig:hrtem}(e) the boundary
between two columnar structures in the AlN:Mn layer is reported. Here, the arrow
\textbf{\textit{a}} indicates a gap between the two columns, while arrow
\textbf{\textit{b}} points to planar defects in the basal plane formed in the
AlN:Mn layer.\\
Having established the lattice parameters (long-range structure), strain state
and Al concentration with XRD, and the microstructure of the layers by means of
TEM, we apply XAS and XES to probe the local atomic and electronic structure
around Mn impurities. The approach employed here follows a well established
method applied in previous studies of GaN:Mn
\cite{Stefanowicz:2010_PRB,Bonanni:2011_PRB}, GaN:Mn,Mg \cite{Devillers:2012_SR}
and related systems as ZnO:Mn \cite{Guda:2013_JAAS} and GaN:Sc
\cite{Knoll:2014_JPCM}. Supported by the complementary spectroscopic techniques
EXAFS, XANES, XLD and XES, we demonstrate that at least 90\% of the Mn atoms
incorporate into the Al$_x$Ga$_{1-x}$N lattice as random substitutional impurities at the
cation site (Mn$_{\rm S}$) with a local spin moment $S$=2 in all the samples
containing up to 82\% of Al.\\
\begin{figure}[!htb]
  \centering
  \includegraphics[width=\MyFigSize]{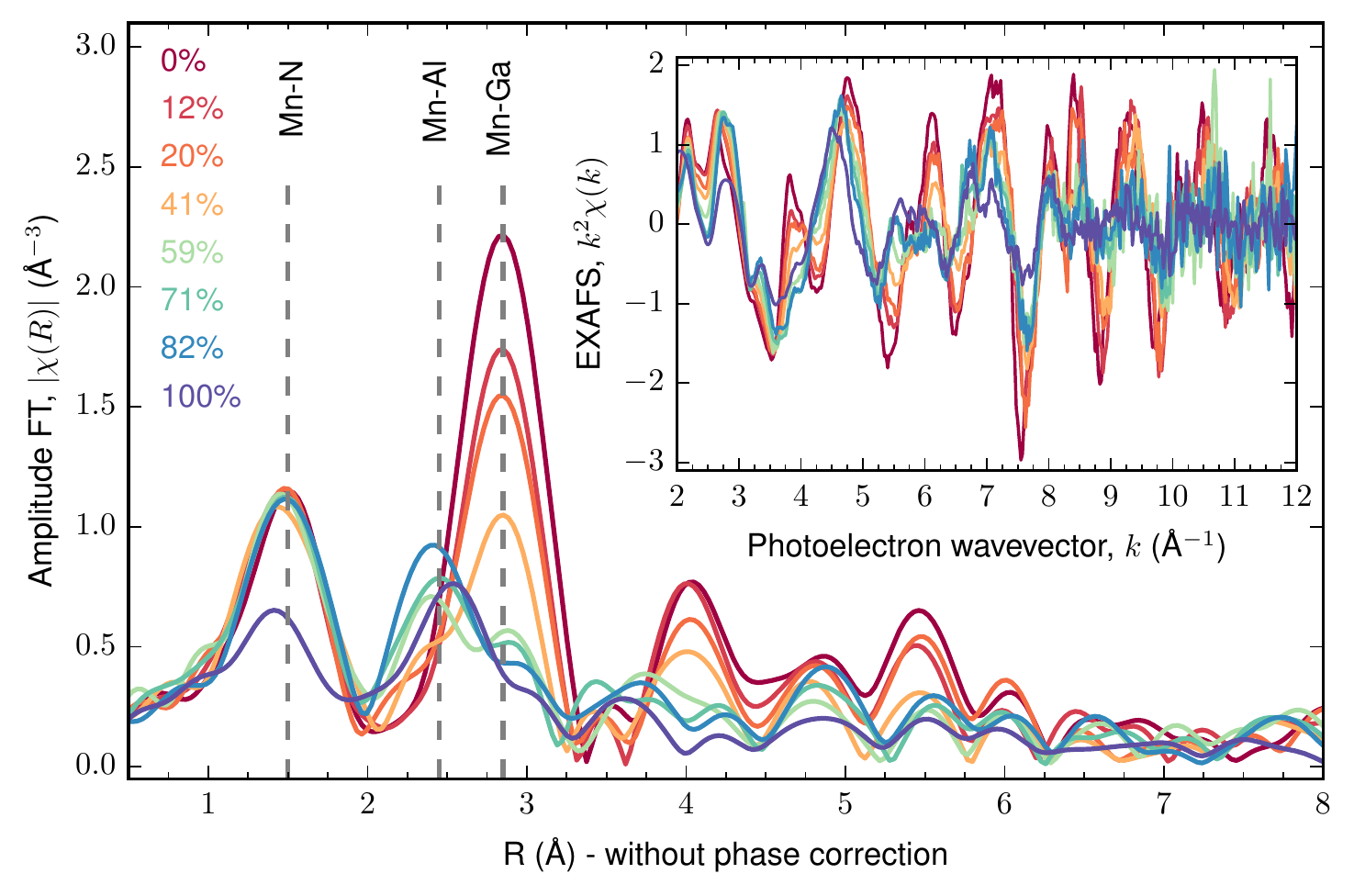}
  \caption{(Color online) Amplitude of the Fourier transform (FT) of the
    $k^2$-weighted EXAFS signal, $\chi(k)$, - shown in the inset - for all the
    samples (identified by Al \%), collected in VGI geometry. The FT is
    performed using an Hanning window ($dk=1$) in the $k$-range [2.5--9.5]
    \AA$^{-1}$. The three vertical dashed lines show the nominal position of the
    Mn-N, Mn-Al and Mn-Ga next nearest neighbors bond distances in R-space
    (without phase correction, that is, the shown R scale does not correspond to
    the absolute bond distances).}
  \label{fig:exafs-ft}
\end{figure}
The EXAFS technique is a well established powerful tool for the local structure
characterization of doped semiconductors
\cite{Boscherini:2008_book,DAcapito:2011_SST}. The system under study is very
challenging for the conventional Fourier transform (FT) quantitative analysis of
the EXAFS data. In fact, not only the lattice is distorted locally by the
introduction of the Mn dopant (similarly to \eg GaN:Mn or AlN:Mn), but also the
alloying effect due to the ternary compound Al$_x$Ga$_{1-x}$N, strongly affects the
resulting spectra that represent an average pair distribution function around
the Mn atoms. As shown in Fig.~\ref{fig:exafs-ft}, there is an evolution of the
EXAFS signal with the Al concentration. The main changes are visible in the
$k$-region [2.5--9.5] \AA$^{-1}$ which is especially sensitive to the Mn next
nearest neighbors average configuration. In particular, the evolution of the
spectral features at $\approx$4\,\AA$^{-1}$ and at $\approx$6\,\AA$^{-1}$ may be
understood by taking into account the destructive interference of the
out-of-phase Mn-Ga and Mn-Al scattering paths in the cation-cation coordination
shells. This effect is evidenced by taking the amplitude of the FT in the range
of interest. The first peak, which represents the Mn-N bond distance, is
substantially constant up to 82\% Al. The second and third main peaks,
corresponding to Mn-Al and Mn-Ga coordination shells, respectively, show a
continuous evolution with increasing Al concentration. In a simple qualitative
analysis and in first approximation, the intensity and position of these peaks
can be ascribed to the coordination number and average bond distance of the
corresponding scattering paths, respectively. The fact that the Mn-N peak is
constant up to 82\% Al points to a Mn$_{\rm S}$ well ordered defect, while the
reduction (increase) in amplitude of the Mn-Ga (Mn-Al) peak is related to the
alloying effect and permits to quantify the local Al concentration and bond
distances. Moreover, the strong overall amplitude reduction for the AlN:Mn
sample (100\% Al) is the hint of a locally disordered environment and is in line
with the disordered micro/nano-structure previously revealed by TEM
measurements.\\
\begin{table*}[!htb]
  \caption{Results of the EXAFS quantitative analysis.}
  \label{tab:exafs-fits}
  \begin{ruledtabular}
    \begin{tabular}{c|ccccccccc}
      sample & $x_{\rm N}$ & $x_{\rm Al}$ & $\sigma^2$ & \multicolumn{2}{c}{$R_{\rm Mn-N}$} & \multicolumn{2}{c}{$R_{\rm Mn-Al}$} & \multicolumn{2}{c}{$R_{\rm Mn-Ga}$} \\
          &    &    &                     & VGI & HGI & VGI & HGI & VGI & HGI \\
          & \% & \% & 10$^{-3}$ \AA$^{-2}$ & \AA & \AA & \AA & \AA & \AA & \AA \\
      \hline
      \#A & 89(9)  & {\em 0}   & 4(2) & 1.99(1) & 1.96(1) & -       & -       & 3.18(1) & 3.18(1) \\ 
      \#B & 80(13) & 10(8)     & 4(2) & 1.97(1) & 1.95(1) & 3.17(1) & 3.18(1) & 3.17(1) & 3.17(1) \\ 
      \#C & 73(10) & 12(8)     & 4(2) & 1.96(1) & 1.94(1) & 3.16(1) & 3.18(1) & 3.16(1) & 3.16(1) \\ 
      \#D & 85(14) & 35(8)     & 8(3) & 1.97(2) & 1.95(2) & 3.21(4) & 3.23(5) & 3.19(5) & 3.19(5) \\ 
      \#E & 85(16) & 54(10)    & 6(3) & 1.95(2) & 1.94(2) & 3.15(5) & 3.18(5) & 3.15(1) & 3.15(7) \\ 
      \#F & 77(18) & 67(9)     & 7(3) & 1.95(1) & 1.94(1) & 3.14(2) & 3.17(3) & 3.13(3) & 3.13(3) \\ 
      \#G & 74(18) & 76(12)    & 7(4) & 1.96(1) & 1.95(1) & 3.12(2) & 3.15(3) & 3.13(3) & 3.13(5) \\ 
      \#H & 63(9)  & {\em 100} & 9(3) & 1.98(1) & 1.98(1) & 3.09(1) & 3.14(1) & -       & -       \\ 
    \end{tabular}
  \end{ruledtabular}
\end{table*}
A quantitative analysis $via$ a least-squares fit of the EXAFS data is then
performed. Due to the complexity of the system under investigation and in order
to keep the correlation between the fitted variables as low as possible, a model
with a minimum set of parameters to describe the whole Al concentration range is
found. This corresponds to the best fitting model and consists of a Mn$_{\rm S}$
defect in Al$_x$Ga$_{1-x}$N\ expanded in three sets of single scattering paths:
Mn-N, Mn-Al and Mn-Ga, corresponding to the first three coordination shells. For
each sample, the fit is performed in R-space, limited to the [1--3.5]
\AA\ range. Both VGI and HGI data sets (weighted by the noise level) are
included in a single fit in order to correctly account for the polarization
effects. This permits to report the average bond distances for the out-of-plane
(VGI, parallel to $c$) and in-plane (HGI, perpendicular to $c$) atomic
configurations. The results are shown in Table~\ref{tab:exafs-fits} and in
Supplementary Fig.~\ref{figS:exafs-fits}. The model is built as follows: the
passive electron reduction factor \cite{Li:1995_PRB}, $S_0^2$, is fixed to the
calculated value of 0.935; the coordination numbers for Mn-N and Mn-Al are
fitted, respectively, $via$ the variables $x_{\rm N}$ and $x_{\rm Al}$, while
the coordination number of the second cation shell is constrained to sum to 12;
a common Debye-Waller factor, which accounts for both the structural and thermal
disorder, is fitted to $\sigma^2$ for all single scattering paths; three
variables are employed for the Mn-N, Mn-Al and Mn-Ga average distances, $R_{\rm
  Mn-N}$, $R_{\rm Mn-Al}$ and $R_{\rm Mn-Ga}$, respectively, with a common
expansion/contraction factor in the two orthogonal directions (VGI and HGI); a
common variable is fitted also for the shift of the edge energy, $\Delta
E_0$. This model permits to keep the numerical correlation between the variables
below a 50\% level. The R-factor of the fits ranges from 0.009 to 0.04,
affecting the propagated error bars, as reported in
Table~\ref{tab:exafs-fits}. Several additional fitting models have been tested,
either increasing the number of fitted variables or introducing additional
scattering paths from other defects, as Mn interstitials (Mn$_{\rm IO}$ and
Mn$_{\rm IT}$). In all cases those models do not pass a F-test
\cite{Michalowicz:1999_JSR,Downward:2007_conf}, meaning that the improvement in
the fit quality is not statistically relevant.\\
\begin{figure}[!htb]
  \centering
  \includegraphics[width=\MyFigSize]{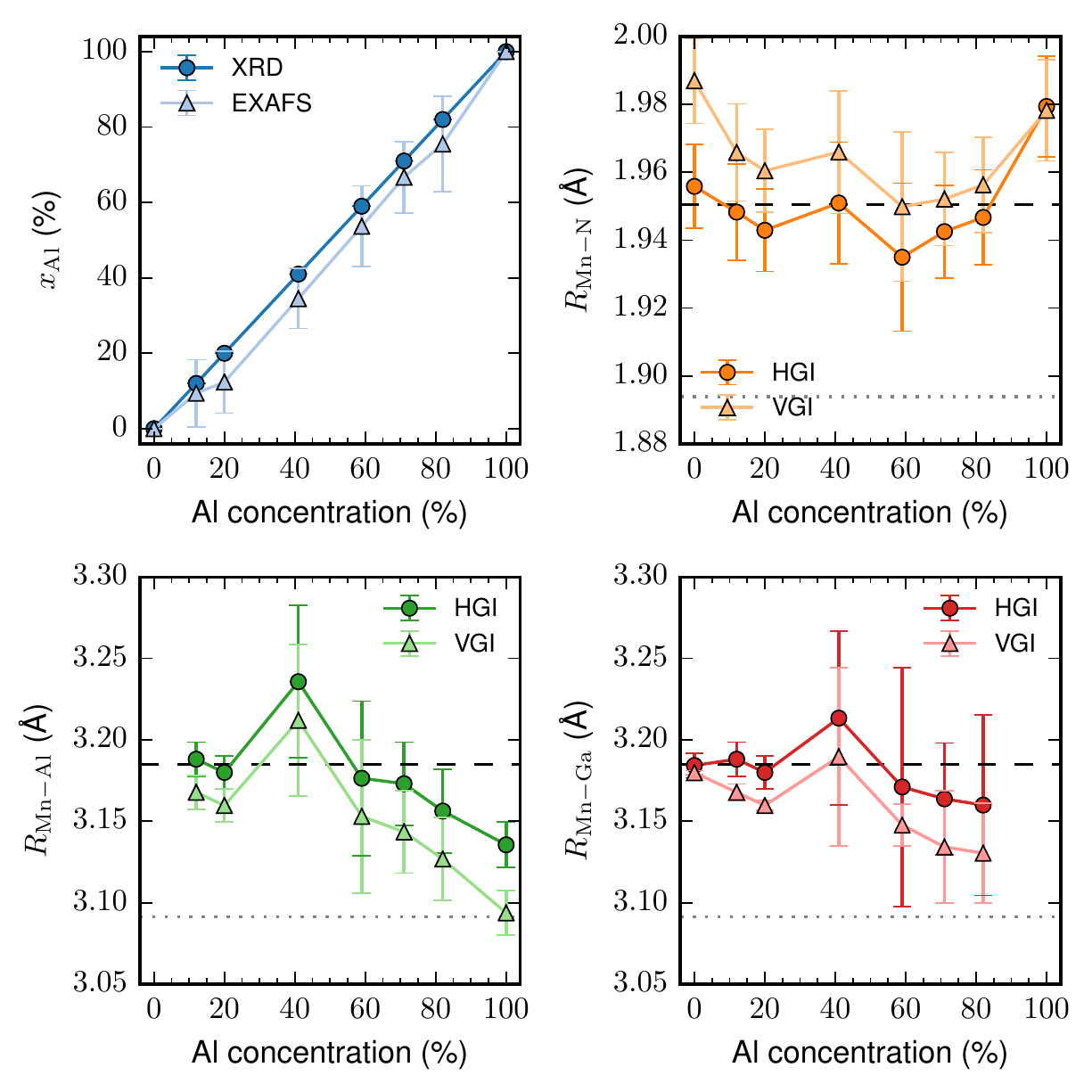}
  \caption{(Color online) Results of the EXAFS quantitative analysis for $x_{\rm
      Al}$, $R_{\rm Mn-N}$, $R_{\rm Mn-Al}$ and $R_{\rm Mn-Ga}$ variables (as
    reported in Table \ref{tab:exafs-fits}). The horizontal lines are the
    corresponding average bond distances for GaN (dashed) and AlN (dotted).}
  \label{fig:exafs-res7}
\end{figure}
The EXAFS quantitative analysis indicates that the majority of Mn atoms is in a
Mn$_{\rm S}$ configuration. On the other hand, the fitted percentage of $x_{\rm
  N}$ does not correspond exactly to the percentage of Mn$_{\rm S}$ in the
samples. In fact, the absolute value of this variable, which represents the
coordination of the first coordination shell (Mn-4N tetrahedron), is affected by
the numerical correlation with $\sigma^2$ and by the presence of nitrogen
vacancies, as found in similar samples \cite{Piskorska-Hommel:2015_JAP}. For
this reason, we rely on the results of the XLD analysis, which is much more
sensitive to the symmetry of the crystal, for determining the level of Mn$_{\rm
  S}$ in the samples. Nevertheless, a strong $k$-independent amplitude reduction
of the EXAFS signal is obtained for the AlN:Mn sample (\#H). As shown by the TEM
micrographs, this sample has a columnar structure, thus the amplitude reduction
is attributed to an increased local disorder, as it was demonstrated by EXAFS
simulations combined with molecular dynamics calculations for Mn nano-columns in
Ge:Mn \cite{Arras:2012_PRB}. The second percentage parameter,$x_{\rm Al}$, is
extracted from the fitted coordination number of the second coordination shell,
keeping the constraint of 12 total neighbors (Ga/Al) dictated by the wurtzite
structure. The results follow a linear dependence and match, within the error
bars, with the Al concentration found by XRD. Furthermore, it is found that the
average Mn-N bond distance is larger than those of Ga-N or Al-N and is not
affected by the Al doping, while Mn-Al and Mn-Ga show a contraction going from
GaN:Mn to AlN:Mn, as expected by the reduction of the lattice parameters. This
implies that the lattice distortion introduced by the Mn incorporation is local
and mainly limited to the first coordination shell.\\
\begin{figure}[!htb]
  \centering
  \includegraphics[width=\MyFigSize]{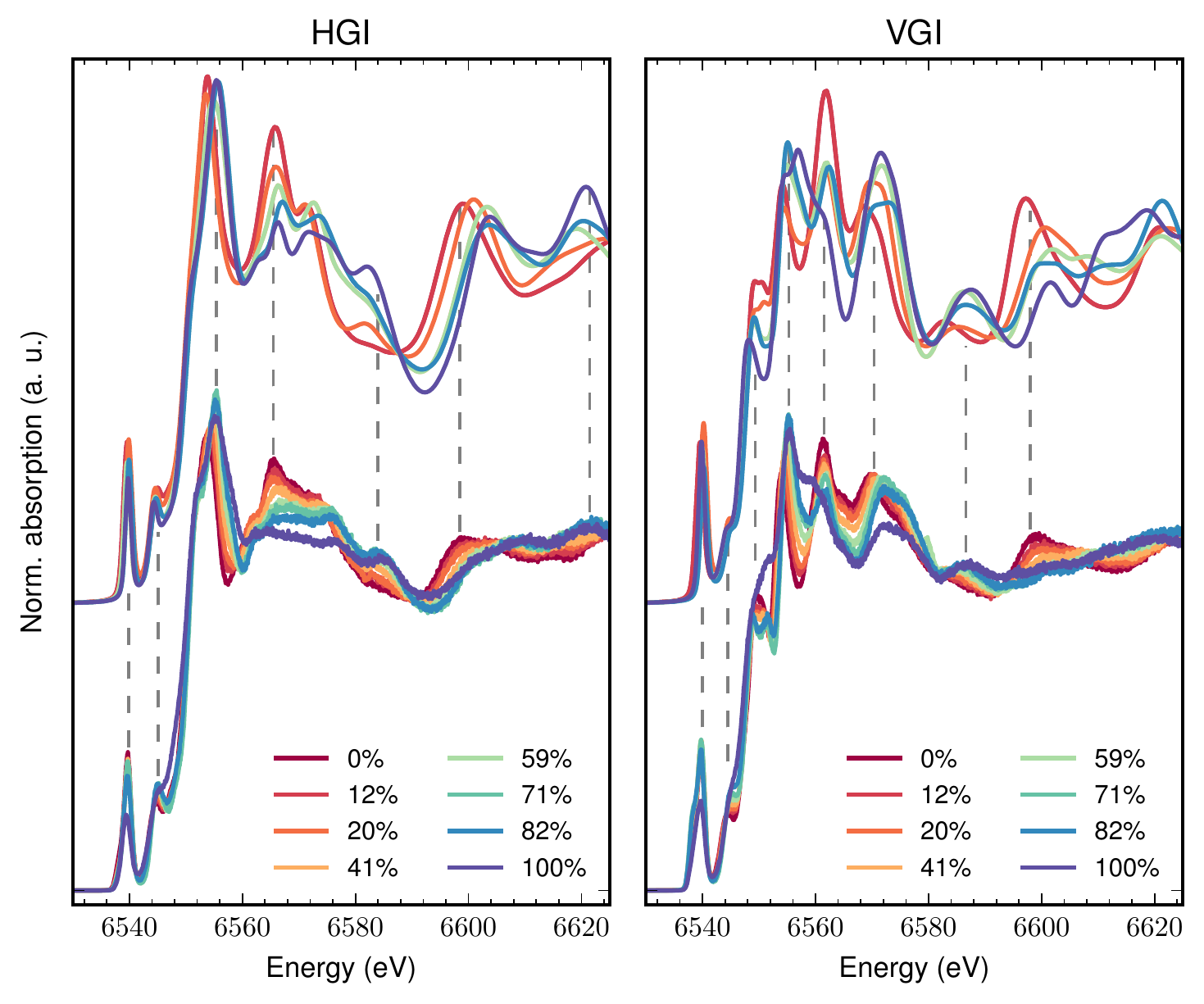}
  \caption{(Color online) Normalized K$\alpha_{1}$ HERFD-XANES spectra (bottom)
    for the Al$_x$Ga$_{1-x}$N:Mn series with the corresponding simulations (top) for HGI
    (left panel) and VGI (right panel) geometries. The vertical dashed lines are
    guides to the eye of the main spectral features.}
  \label{fig:xanes}
\end{figure}
In order to further confirm the local structural description obtained $via$
EXAFS analysis, the XANES region is investigated through {\em ab initio}
simulations. In Fig.~\ref{fig:xanes} the normalized K$\alpha_1$ HERFD-XANES
spectra are shown together with their relative simulations (using the {\sc
  fdmnes} \cite{Bunau:2009_JPCM} code) for the HGI and VGI geometries. The
HERFD-XANES spectra correspond to a diagonal cut in the 2D resonant inelastic
x-ray scattering (RIXS) plane \cite{Rovezzi:2014_SST} and can be approximated to
a standard XANES spectra only in the region above the main absorption edge,
where the spectral features arise from electric dipole transitions from 1$s$ to
4$p$ empty states of the absorbing atoms (Mn). On a first order approximation,
this energy range can be described by multiple scattering theory employing
simple muffin-tin potentials \cite{Slater:1937_PR} within a one-electron
approach, \ie the level of theory employed for the simulated spectra shown in
this study. The spectral features present in the pre-edge region of the
HERFD-XANES spectra cannot be fully described by the level of theory employed
here and a quantitative analysis requires to account for the full RIXS plane,
not only for line cuts \cite{Rovezzi:2014_SST}. Nevertheless, the presence of an
intense pre-edge peak in the K-edge XAS spectra of 3$d$ transition metals is the
fingerprint of tetrahedral (T$_{\rm D}$) symmetry
\cite{Westre:1997_JACS,Yamamoto:2008_XRS}, due to allowed electric dipole
transitions to the $p$-character of the $t_2$ spin-polarized 3$d$ states. The
spectral features present in the XANES region do not correlate straightforward
with a given coordination shell or scattering species, but are the result of
full multiple scattering configurations. This induces an enhanced sensitivity to
the geometry around the absorber. On the other hand, this also makes challenging
to quantitatively model the XANES via {\em ab initio} methods. As shown in
Fig.~\ref{fig:xanes}, all the spectral features and the trend with increasing Al
concentration are reproduced by the simulations using a substitutional model
based on the DFT-relaxed supercells, rescaled to the experimental lattice
parameters. To better evaluate the quality of each simulation, the supplementary
Figs.~\ref{figS:xanes-subs} and \ref{figS:xanes-ints} show the comparison with
experimental spectra for the nominal Wyckoff sites and the DFT-relaxed
positions. The defects investigated are Mn$_{\rm S}$, Mn$_{\rm IT}$ and Mn$_{\rm
  IO}$ in Al$_x$Ga$_{1-x}$N. In order to get more quantitative results, a linear
combination fit (LCF) analysis of the XANES spectra is performed. The
constraints imposed are: the presence of the Mn$_{\rm S}$ phase; the number of
components is limited to two (one substitutional and one interstitial); an
energy shift for the interstitial phase is allowed (fitted). All combinations
among the four interstitial cells are performed and the fits are ranked by
$\chi^2$. In all samples/geometries it is found that the Mn$_{\rm S}$ phase is
$>$80\% and the complementary phase is the non relaxed Mn$_{\rm IT}$ defect. On
the other hand, the $\chi^2$ values of the best fits do not pass a statistical
test (F-test), meaning that the increase in the fit quality is not
relevant. This confirms what previously found by EXAFS and the formation
energies results of the DFT, that is, that Mn$_{\rm IT}$ defect in
Al$_x$Ga$_{1-x}$N is not stable and has a high formation energy.\\
A more quantitative analysis to establish the percentage of Mn atoms
incorporating as substitutional defects in the host matrix, is obtained by
studying the XLD spectra. It is established that XLD is extremely sensitive to
the symmetry of non-cubic sites \cite{Brouder:1990_JPCM} and it was shown to be
a po\-wer\-ful tool to determine the quality of substutional inclusions in
dilute magnetic semiconductors \cite{Ney:2010_NJP}. The XLD spectra for the
studied samples are reported in Fig.~\ref{fig:xld} and are obtained from the
difference between the HERFD-XANES spectra in VGI and HGI geometries. The
amplitude of the XLD main oscillation at the edge position highlighted in
Fig.~\ref{fig:xld} is taken as a figure of merit for Mn$_{\rm S}$. In fact, the
maximum XLD amplitude would be obtained for 100\% Mn$_{\rm S}$ dilute in a
perfect Al$_x$Ga$_{1-x}$N lattice. The Mn$_{\rm IT}$ interstitial shows a XLD
signal too, however it is not in phase with the Mn$_{\rm S}$ XLD signal and the
resulting XLD amplitude in the region of interest is reduced. As a reference for
the 100\% case, we arbitrarily rescale the experimental XLD amplitudes to the
XLD amplitude at the Ga K-edge of a GaN:Mn layer from
Ref.~\cite{Wilhelm:2007_conf}. The results are reported in the inset to
Fig.~\ref{fig:xld}. The increasing values of the Mn$_{\rm S}$ percentage for Al
$\le$82\% are due to the accuracy of the normalization procedure employed. In
fact, a more accurate procedure would require to rescale the Mn K-edge XLD
amplitudes to the Ga K-edge (or Al K-edge) XLD amplitude measured for each
sample in the same experimental conditions. On the other hand, the systematic
errors are estimated to be within a $\pm$10\% bandwidth. The dramatically low
Mn$_{\rm S}$ value for AlN:Mn can be safely attributed to an actual reduction of
Mn$_{\rm S}$ in this sample.\\
As final point we discuss the Mn valence state inferred from the integral of the
absolute difference of the K$\beta$ XES data (integrated absolute difference --
IAD analysis \cite{Rovezzi:2014_SST}). This method is preferred over the one
employing the position of the main absorption edge for the possibility it gives
to quantitatively follow the evolution of the effective spin moment on Mn
($S_{\rm eff}$) as a function of a given parameter and to directly compare the
results with DFT calculations \cite{Rovezzi:2014_SST}. The total magnetic moment
per unit cell calculated with DFT is in all cases 4\,$\mu_{B}$ and corresponds
to $S_{\rm eff}$\,$\approx$\,2.0, as found in the frame of a Bader partitioning
scheme \cite{Rovezzi:2014_SST}. This result is confirmed experimentally, as
reported in Fig.~\ref{fig:xes}. The Mn valence state is constant within the error
bar for the whole series, with the exception of the AlN:Mn sample, as expected
and supporting all previous results.
\section{Conclusions and Outlook}
\label{sec:outlook}
%
We have carried out an extensive study of epitaxial Al$_x$Ga$_{1-x}$N:Mn on a
series of samples with Al concentration up to 100\%. By XRD we have found that
the Al content in the layers matches -- over the sample series -- the one
expected from growth conditions. The lattice parameters as a function of the Al
concentration are also obtained by XRD. The DFT computations on the formation
energy for the incorporation of Al in a GaN matrix let us to conclude that Al
and Ga are randomly distributed into the lattice, and in Al$_x$Ga$_{1-x}$N:Mn
the Mn ions have the tendency to preferentially substitute for Ga. The formation
of Mn interstitial defects is not favored. A coherent growth without local
aggregation or precipitation is obtained for Al$_x$Ga$_{1-x}$N:Mn with Al
concentrations up to 82\%, confirming the surfactant role of Mn already reported
\cite{Devillers:2015_CGD}. Synchrotron radiation XAS has been employed to probe
the local atomic and electronic structure of Mn. From EXAFS, XANES and XLD it is
found that the majority of the Mn ions is dilute, \ie homogeneously distributed
over the doped layers. An IAD analysis of the XES data allows to determine the
valence state of Mn as constant up to an Al concentration of 82\%. Due to the
reduced lattice parameters with respect to \eg GaN:Mn, enhanced hybridization of
the orbitals can be expected in Al$_x$Ga$_{1-x}$N:Mn, making it a material
system worth to be investigated in view of spintronic functionalities. Moreover,
this work paves the way to the understanding and control of the role played by
Mn in particular and transition metals in general on the structure and
properties of the alloys Al$_x$Ga$_{1-x}$N:TM. Significantly, the incorporation
of Mn has been found to promote the growth of Al$_x$Ga$_{1-x}$N on GaN, to defer
the relaxation of the layers and to increase the critical thickness also for Al
concentrations up to 82\%, with remarkable potential effects on the fabrication
of \eg distributed Bragg mirrors for III-nitride-based optoelectronic devices.\\
%
\begin{acknowledgments}
\section{Acknowledgments}
The authors gratefully acknowledge the European Synchrotron Radiation Facility
(ESRF) for providing synchrotron radiation beam-time (proposal HE-3825). This
work was supported by the ESRF trainee program, by the Austrian Science
Fundation (FWF Projects 24471 and 26830), by the NATO Science for Peace
Programme (Project 984735), by the EU 7$^{\rm th}$ Framework Programme through
the CAPACITIES project REGPOT-CT-2013-316014 and by the European Research
Council (Advanced Grant 22790).\\
\end{acknowledgments}

\bibliographystyle{apsrev4-1}
%


\appendix
\pagebreak
\clearpage
\begin{center}
\textbf{\Large Supplemental Information}
\end{center}
\setcounter{equation}{0}
\setcounter{figure}{0}
\setcounter{table}{0}
\setcounter{page}{1}
\makeatletter
\renewcommand{\theequation}{S\arabic{equation}}
\renewcommand{\thefigure}{S\arabic{figure}}
\renewcommand{\thetable}{S\arabic{table}}
\renewcommand*{\thepage}{S\arabic{page}}
\renewcommand{\bibnumfmt}[1]{[S#1]}
\renewcommand{\citenumfont}[1]{S#1}
%
\section{EXAFS}
\begin{figure*}[!htb]
  \centering
  \includegraphics[width=\MyFigSize]{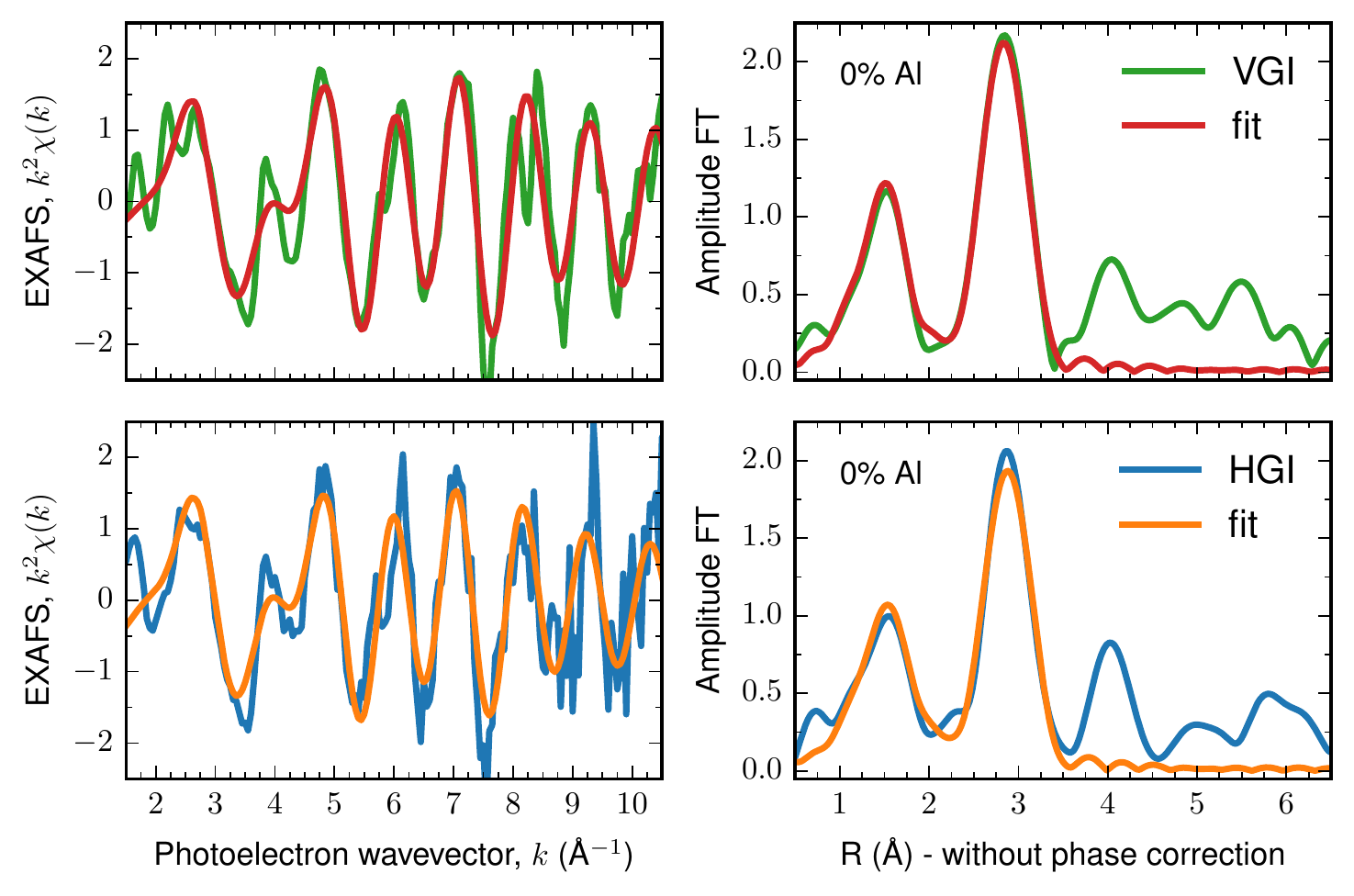}
  \includegraphics[width=\MyFigSize]{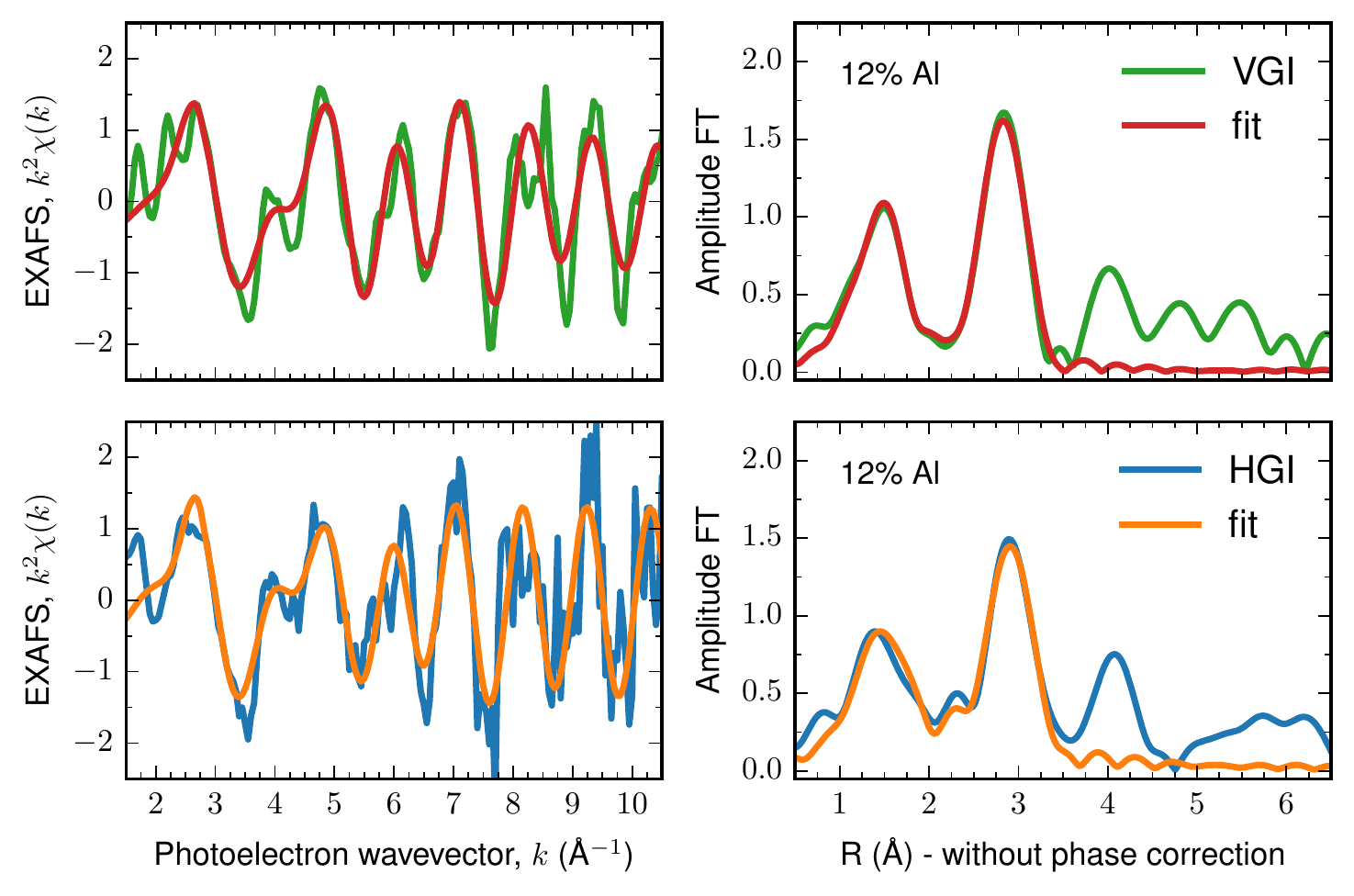}\\
  \includegraphics[width=\MyFigSize]{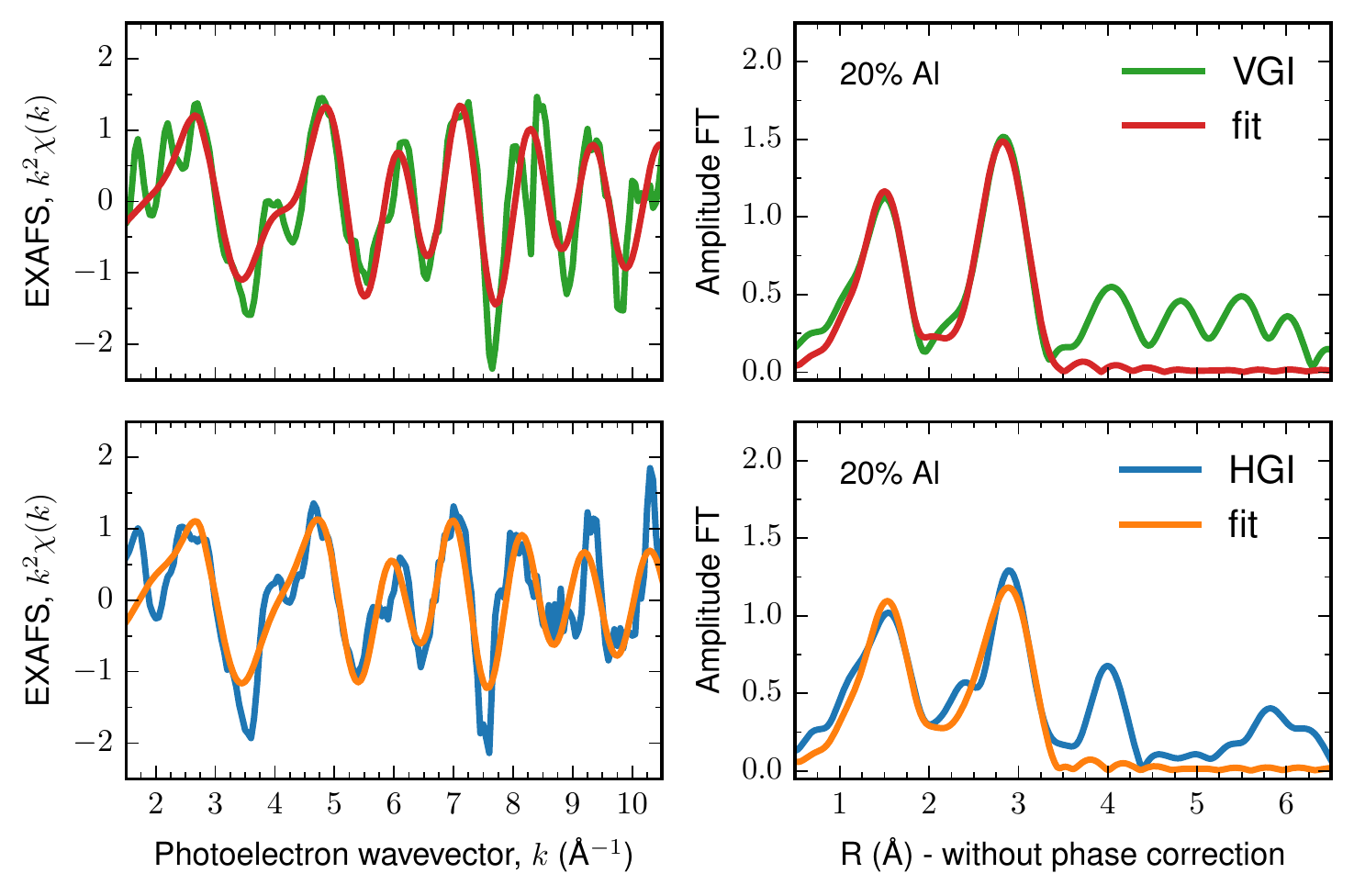}
  \includegraphics[width=\MyFigSize]{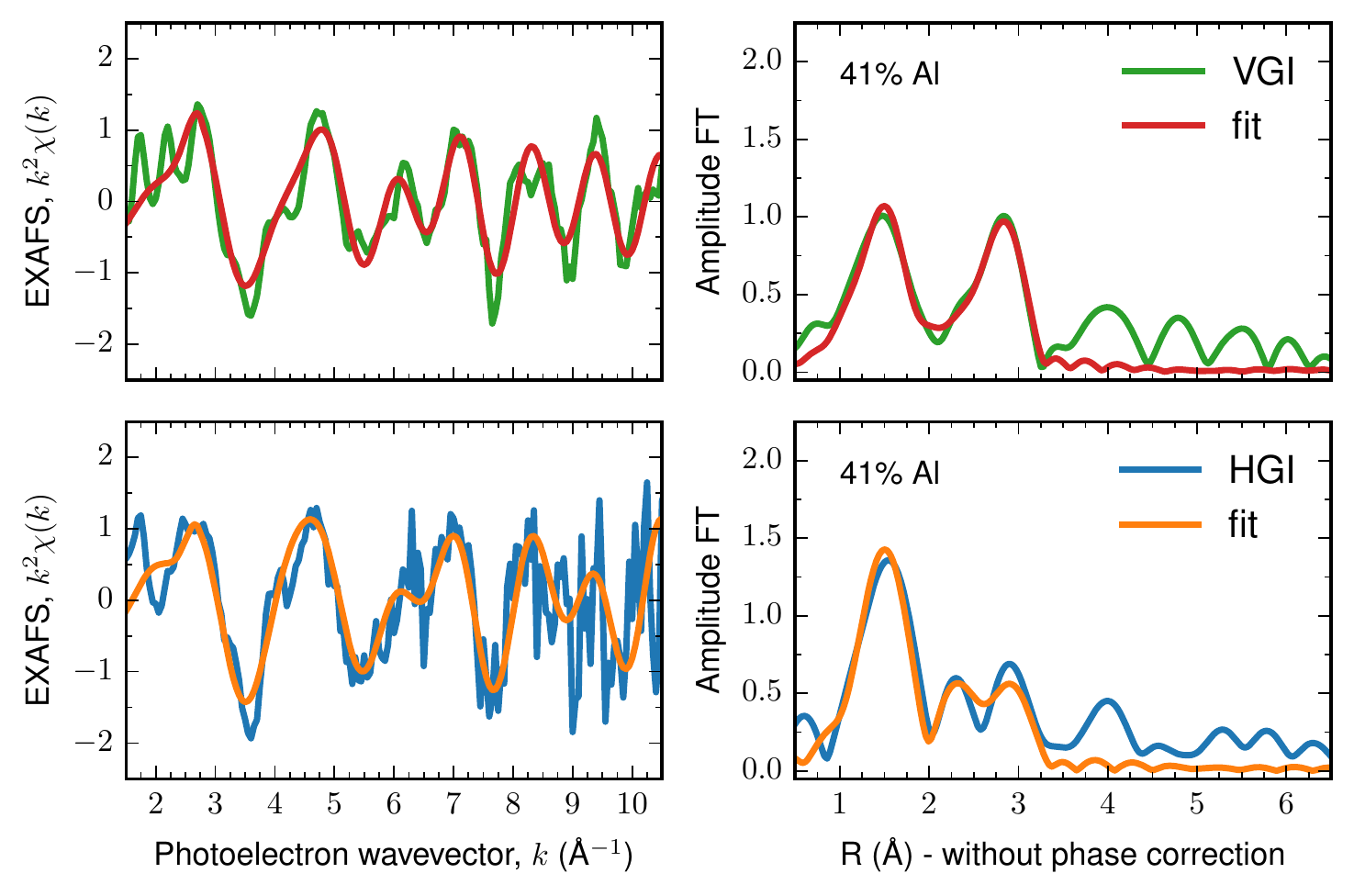}\\
  \includegraphics[width=\MyFigSize]{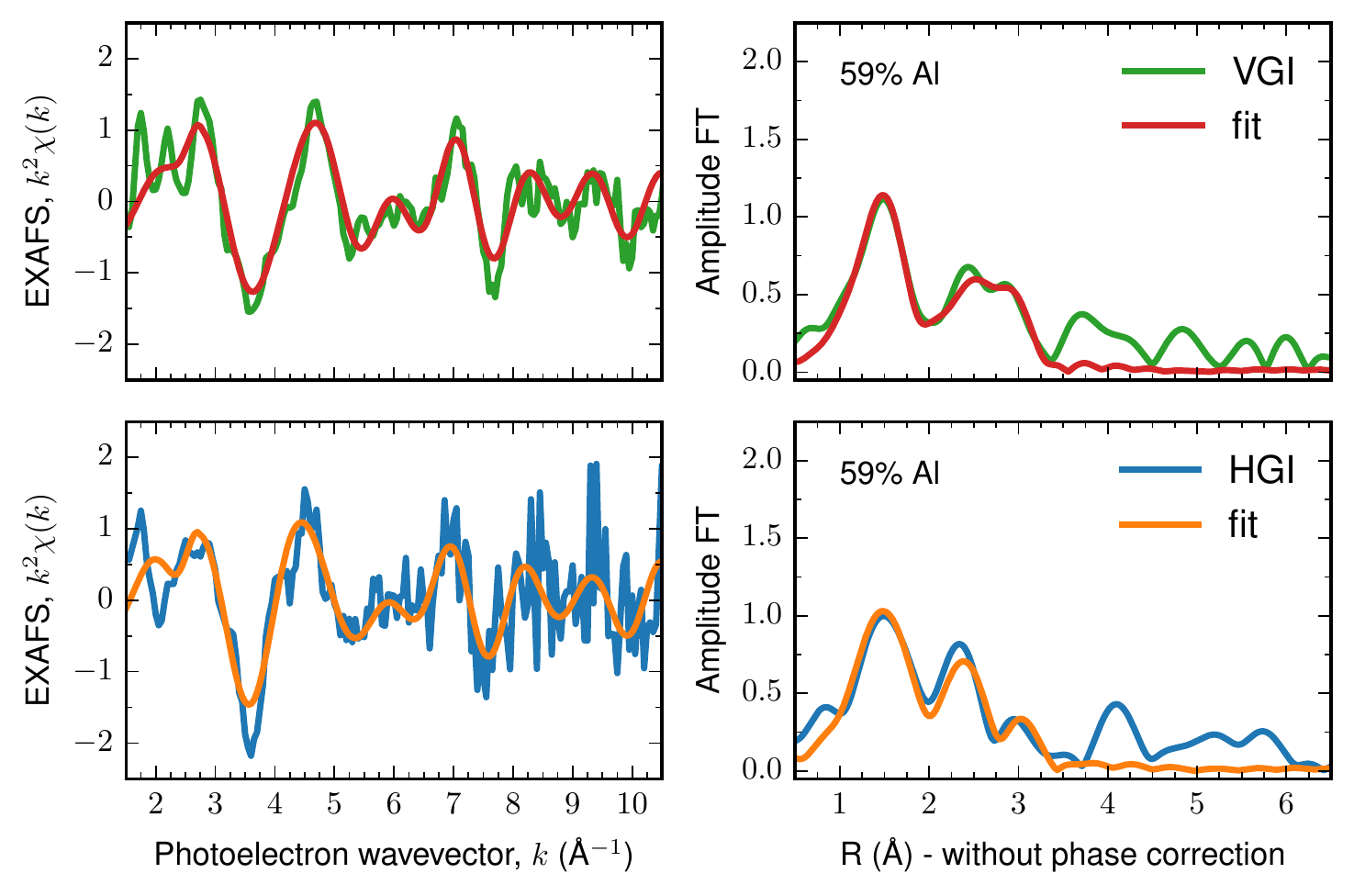}
  \includegraphics[width=\MyFigSize]{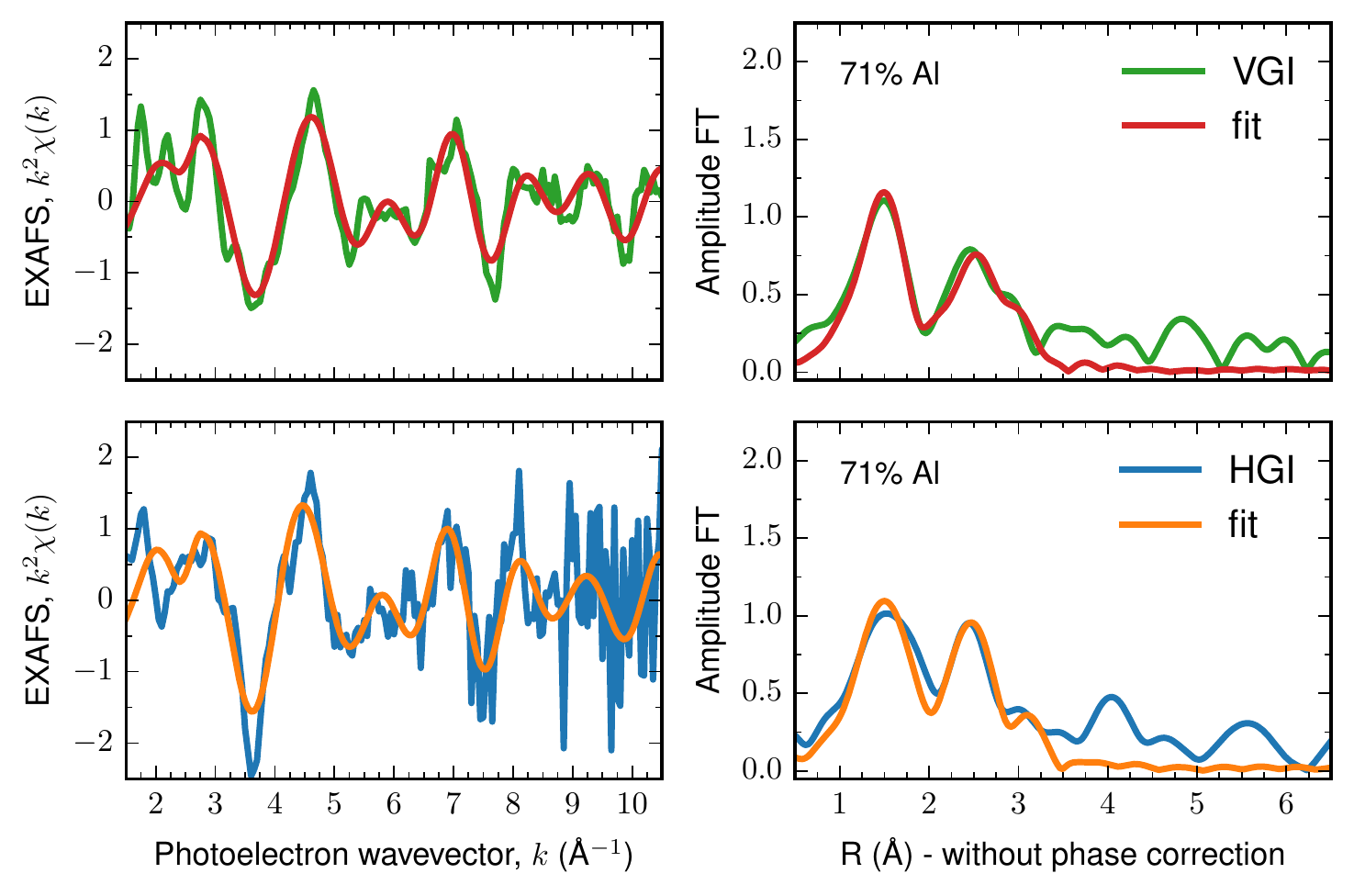}\\
  \includegraphics[width=\MyFigSize]{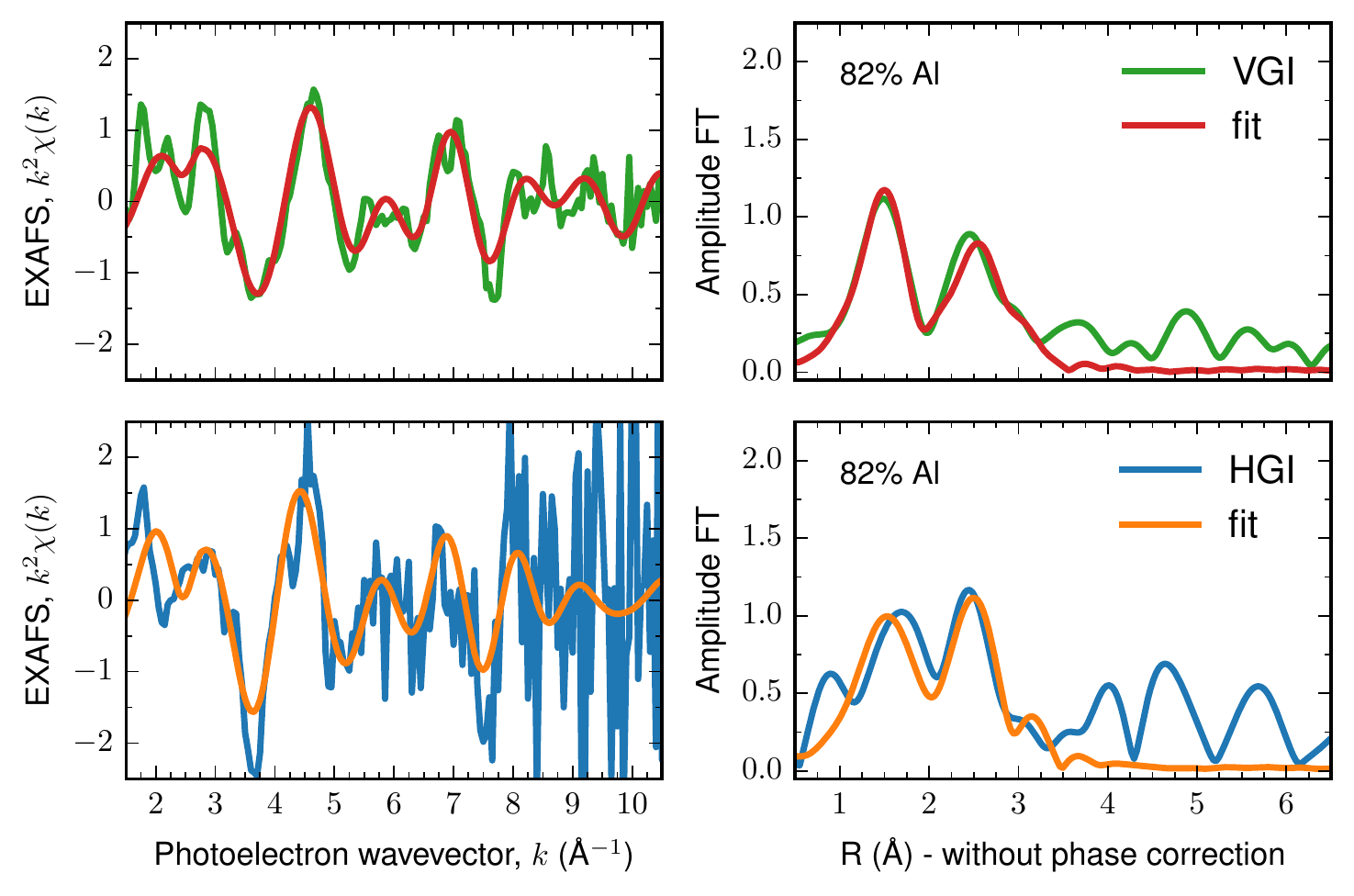}
  \includegraphics[width=\MyFigSize]{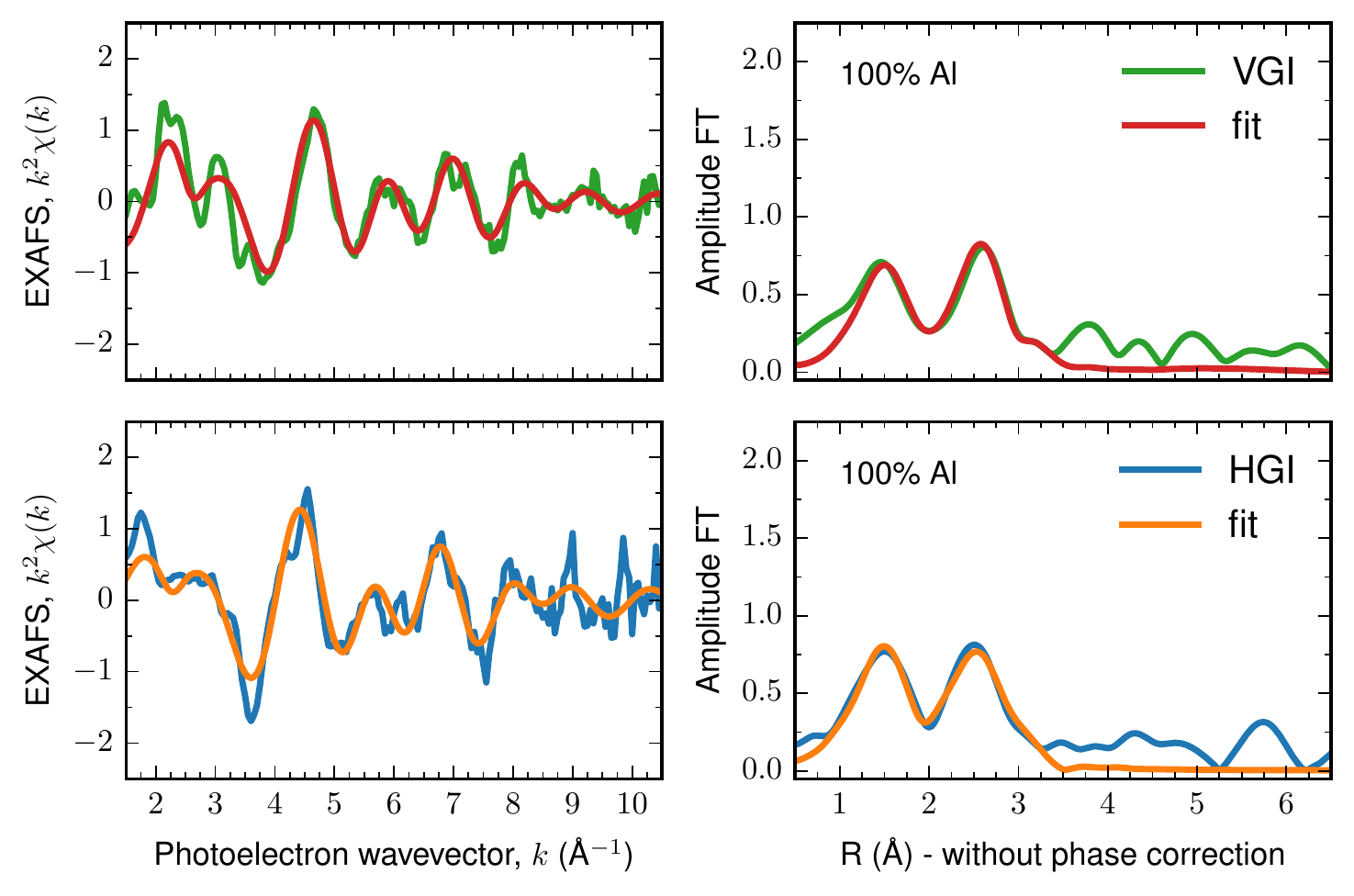}
  \caption{(Color online) EXAFS fits for all samples.}
  \label{figS:exafs-fits}
\end{figure*}

In the plots of Fig.~\ref{figS:exafs-fits}, the quality of the fits performed on
the EXAFS data for both VGI and HGI geometries is reported.
%
\section{XANES}
\begin{figure*}[!htb]
  \centering
  \includegraphics[width=0.75\textwidth]{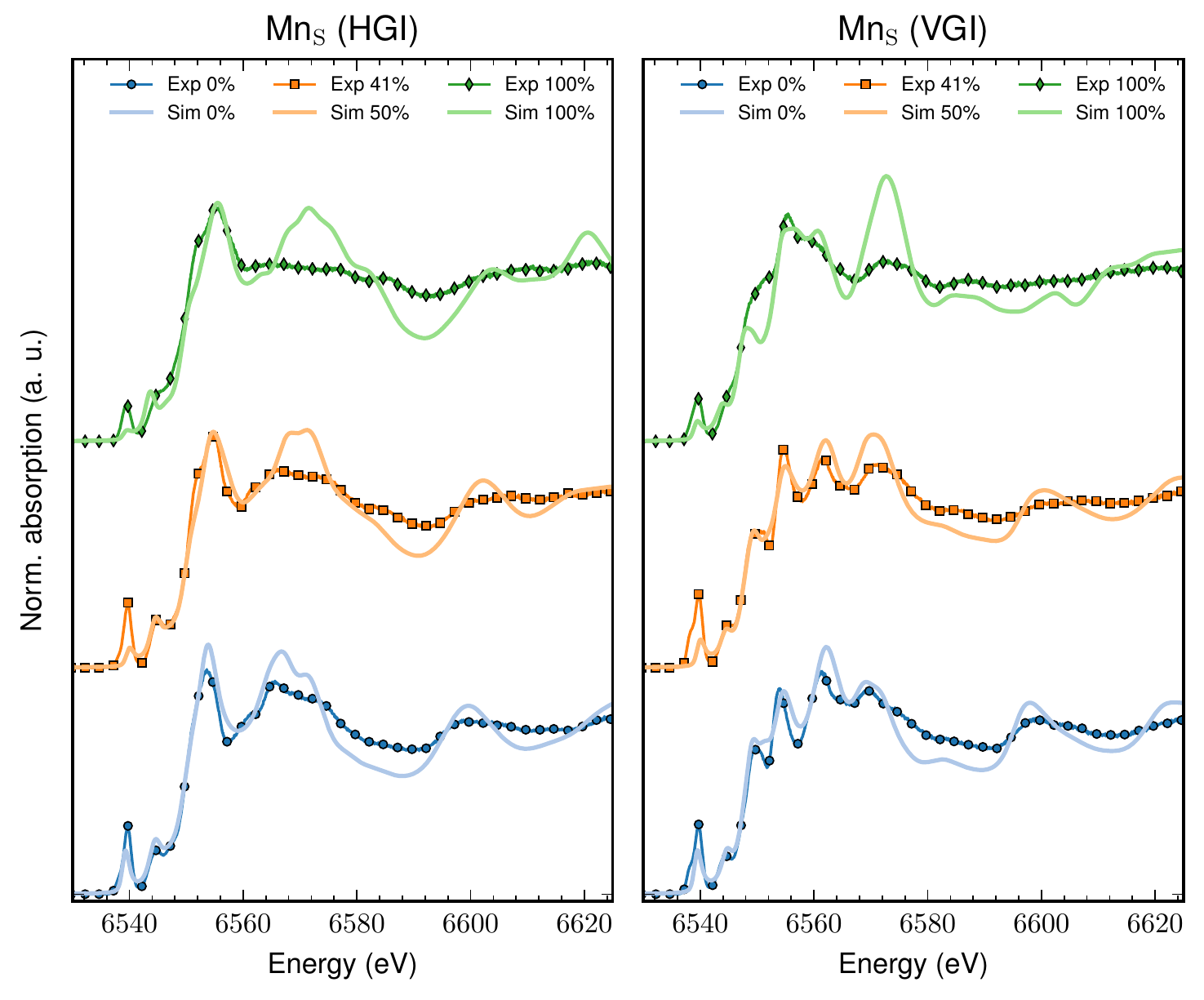}\\
  \includegraphics[width=0.75\textwidth]{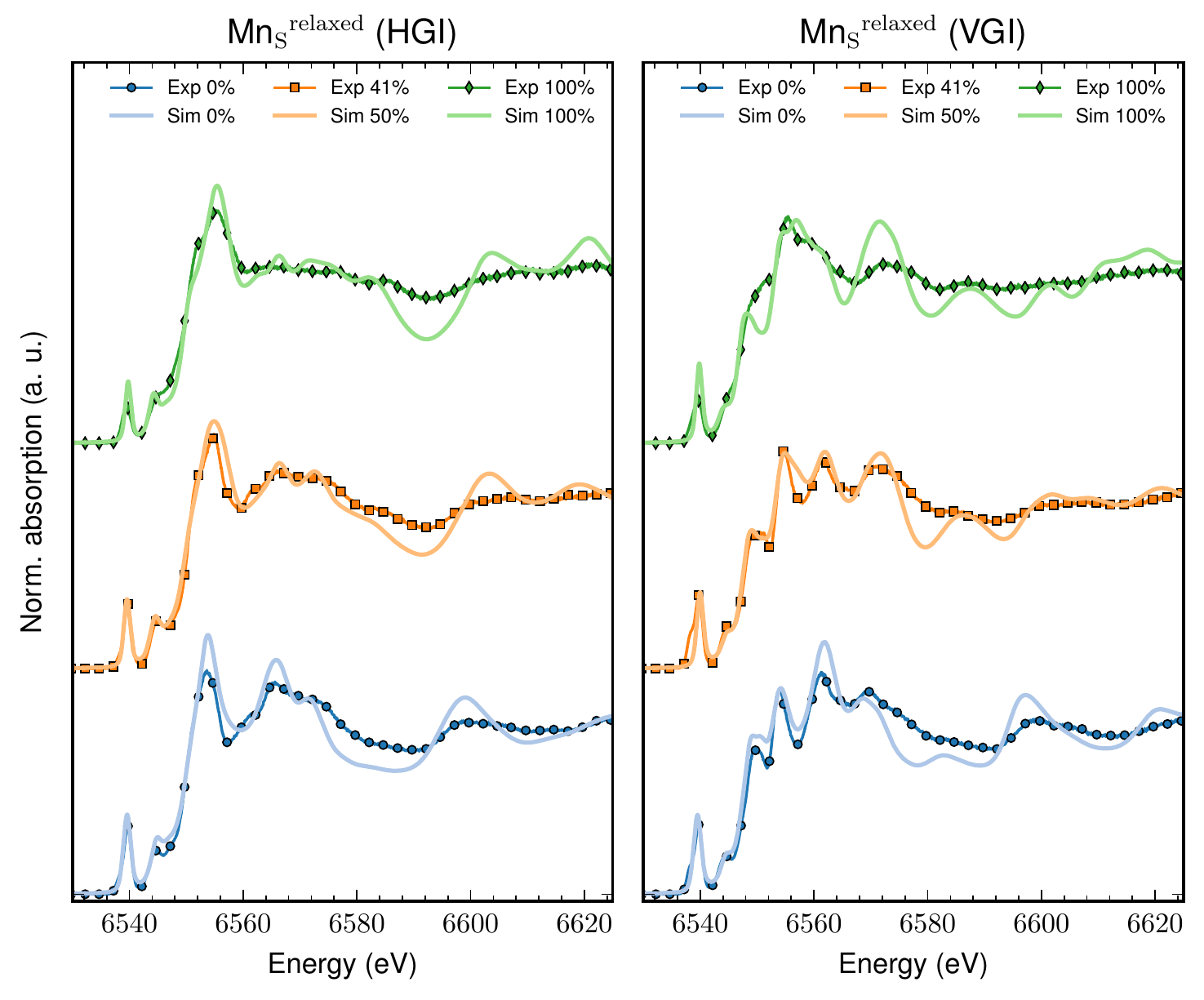}\\
  \caption{(Color online) Simulated XANES spectra in HGI and VGI geometries: for
    Mn$_{\rm S}$ substitutional for nominal and DFT-relaxed supercells. For
    clarity, only three representative Al compositions are shown in the plots:
    0\% (0\%), 41\% (50\%) and 100\% (100\%) from experiment (simulation).}
  \label{figS:xanes-subs}
\end{figure*}

\begin{figure*}[!htb]
  \centering
  \includegraphics[width=0.49\textwidth]{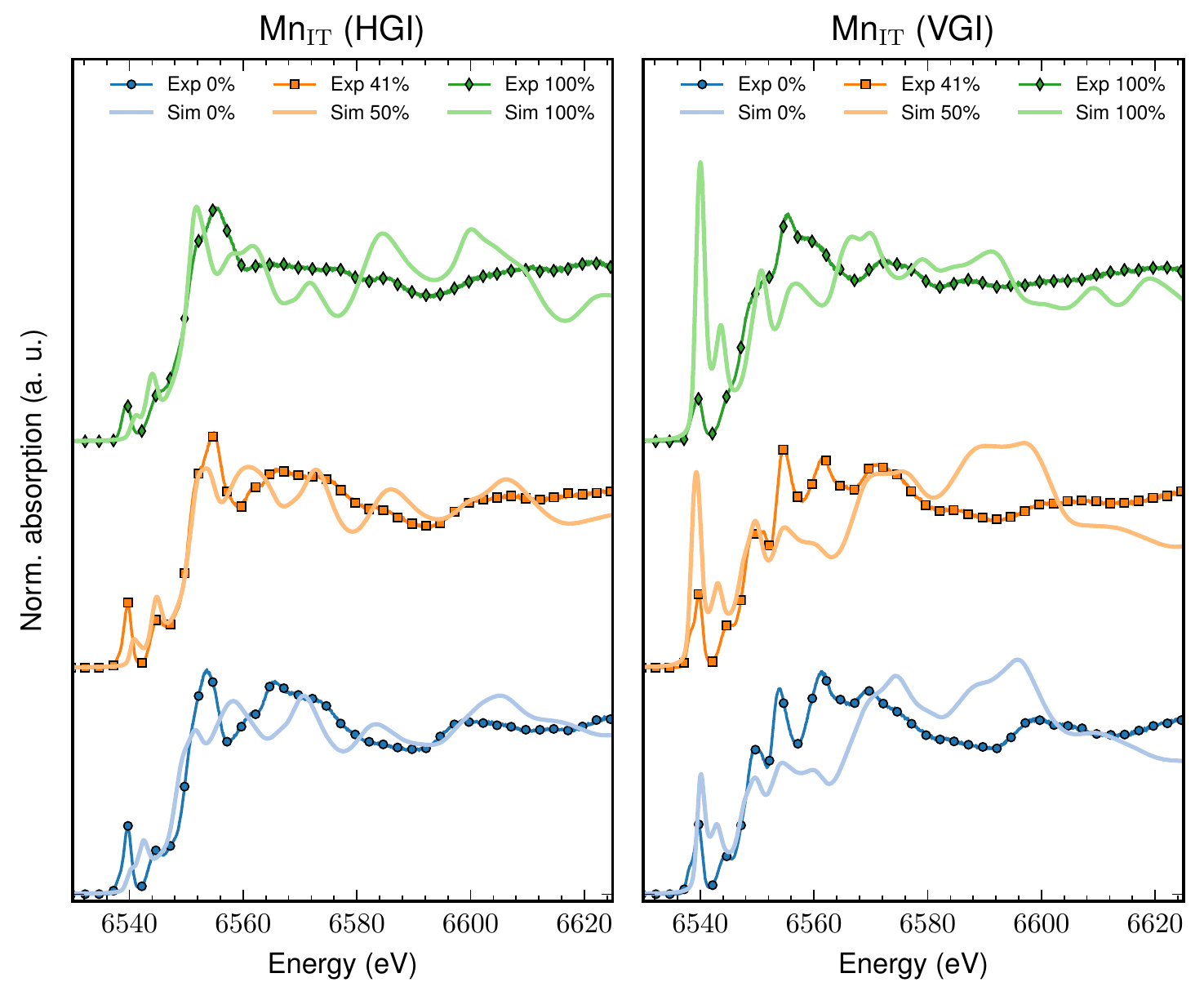}
  \includegraphics[width=0.49\textwidth]{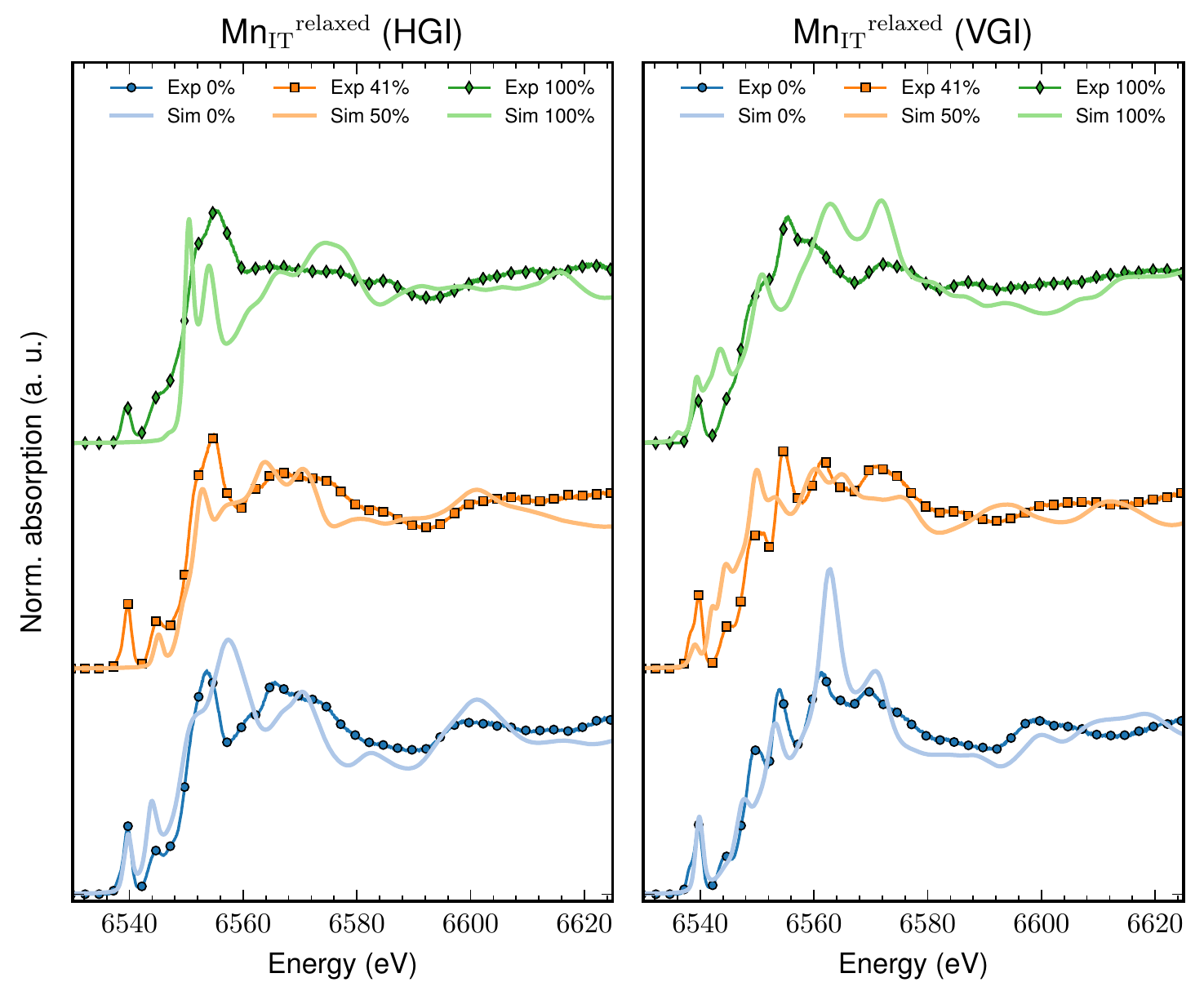}\\
  \includegraphics[width=0.49\textwidth]{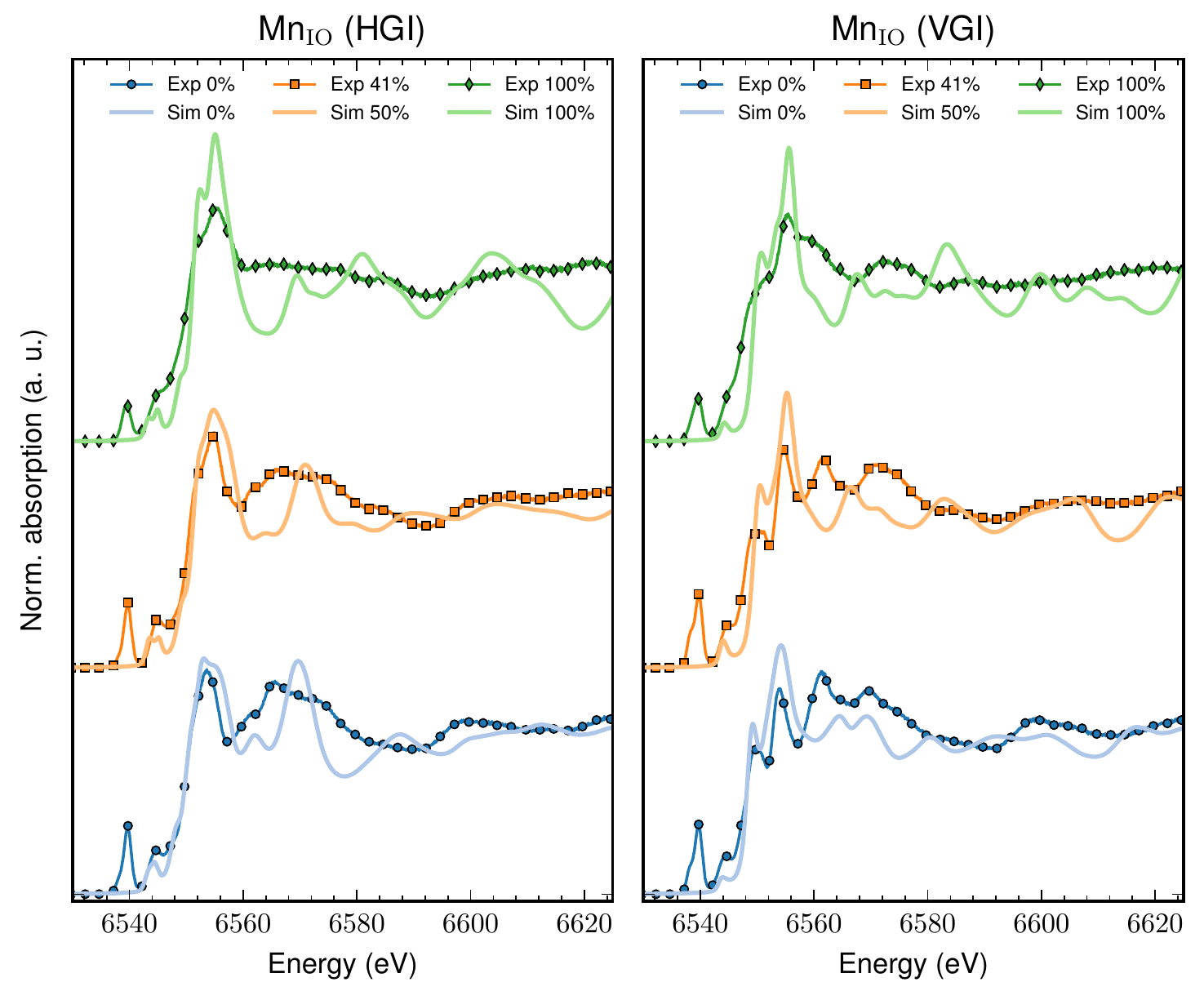}
  \includegraphics[width=0.49\textwidth]{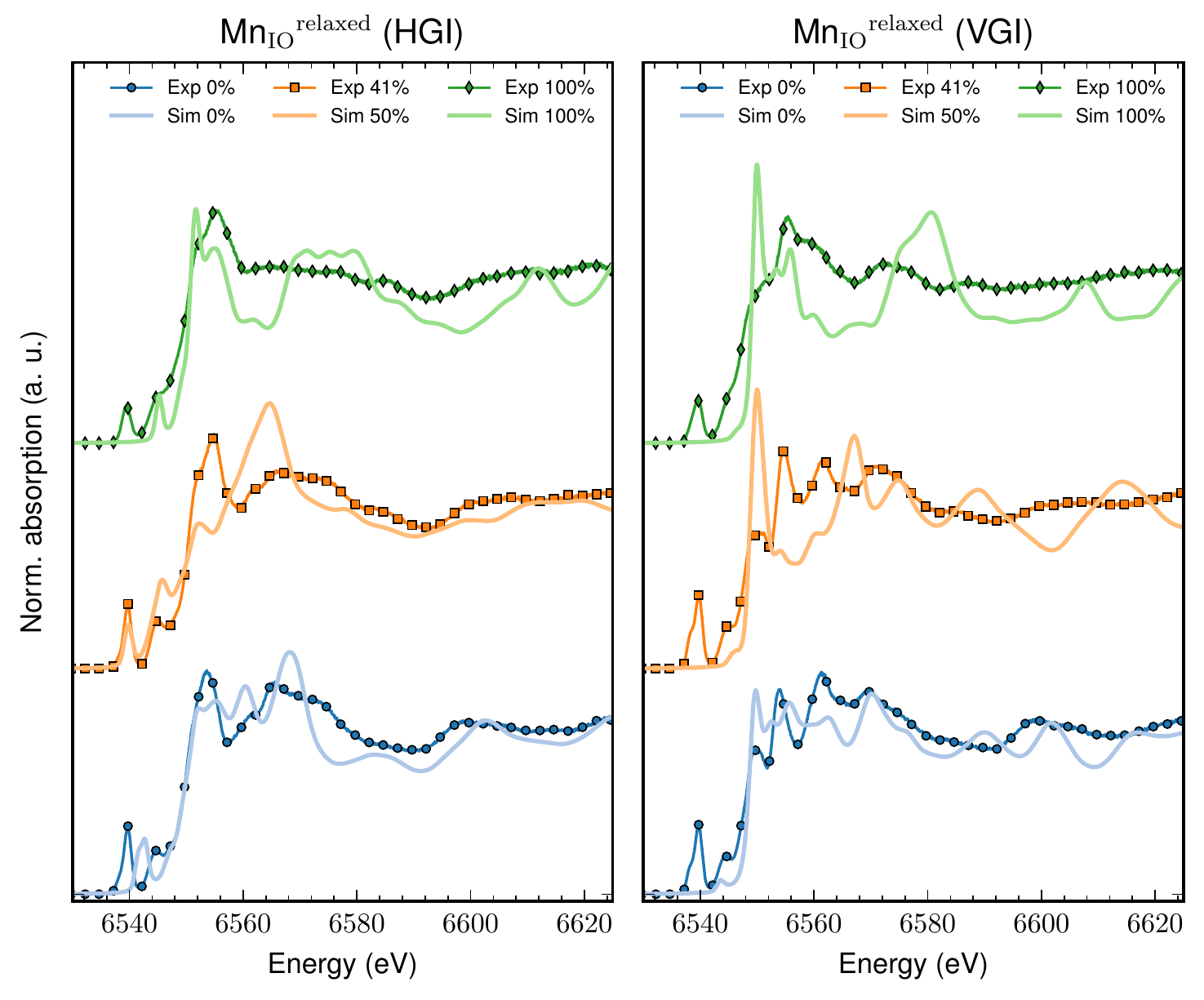}\\
  \caption{(Color online) Simulated XANES spectra in HGI and VGI geometries: for
    Mn$_{\rm IT}$ and Mn$_{\rm IO}$ interstitials for nominal and DFT-relaxed
    supercells. For clarity, only three representative Al compositions are shown
    in the plots: 0\% (0\%), 41\% (50\%) and 100\% (100\%) from experiment
    (simulation).}
  \label{figS:xanes-ints}
\end{figure*}

The quality of the simulated XANES spectra for Mn$_{\rm S}$ is shown in
Fig.~\ref{figS:xanes-subs}, while the quality of the simulated XANES for the
Mn$_{\rm IT}$ and Mn$_{\rm IO}$ interstitials in Al$_x$Ga$_{1-x}$N is given in
Fig.~\ref{figS:xanes-ints}.
%


\end{document}